\documentclass[12pt,letter]{article}

\usepackage{subfigure}
\usepackage{graphicx, epsfig, color}
\usepackage{amsmath,amssymb}
\def\to{\rightarrow}

\def\bi{\begin{itemize}}
\def\ei{\end{itemize}}

\def\tG{\tilde G}

\def\sps1ap{SPS1a$^\prime$}
\def\c1p{C1$^\prime$}

\def\tst{\tilde t}

\def\tg{\tilde g}
\def\tnu{\tilde\nu}

\def\tw{\widetilde W}
\def\tz{\widetilde Z}
\def\alt{\lesssim}
\def\agt{\gtrsim}
\def\be{\begin{equation}}  
\def\ee{\end{equation}}  
\def\bea{\begin{eqnarray}}  
\def\eea{\end{eqnarray}}  
\def\beas{\begin{eqnarray*}}  
\def\eeas{\end{eqnarray*}}

\begin{document}
\begin{titlepage}
\begin{flushright}
OU-HEP-190204 \\
CTPU-PTC-19-06
\end{flushright}

\vspace{0.1cm}
\begin{center}
{\Large \bf Revisiting the SUSY $\mu$ problem\\ 
and its solutions in the LHC era
}\\ 
\vspace{.5cm} \renewcommand{\thefootnote}{\fnsymbol{footnote}}
{\large Kyu Jung Bae$^{1}$\footnote[1]{Email: kyujungbae@ibs.re.kr},
Howard Baer$^{2}$\footnote[2]{Email: baer@nhn.ou.edu },
Vernon Barger$^{3}$\footnote[3]{Email: barger@pheno.wisc.edu }
and Dibyashree Sengupta$^{2}$\footnote[4]{Email: Dibyashree.Sengupta-1@ou.edu}
%and Hasan Serce$^{3}$\footnote[4]{Email: serce@ou.edu } 
}\\ 
\vspace{.5cm} \renewcommand{\thefootnote}{\arabic{footnote}}
{\it 
$^1$Center for Theoretical Physics of the Universe,\\
Institute for Basic Science (IBS), 
Daejeon 34126, Korea\\
}
{\it 
$^2$Homer L. Dodge Dep't of Physics and Astronomy,
University of Oklahoma, Norman, OK 73019, USA \\
}
{\it 
$^3$Dep't of Physics,
University of Wisconsin, Madison, WI 53706, USA \\
}
\end{center}

\vspace{0.3cm}
\begin{abstract}
\noindent 
The supersymmetry preserving $\mu$ parameter in SUSY theories is naively expected
to be of order the Planck scale while phenomenology requires it to be 
of order the weak scale.
This is the famous SUSY $\mu$ problem. 
Its solution involves two steps: 1. first forbid $\mu$, perhaps via some symmetry, 
and then 2. re-generate it of order the scale of soft SUSY breaking terms. 
However, present LHC limits suggest the soft breaking scale $m_{soft}$ 
lies in the multi-TeV regime whilst naturalness
requires $\mu\sim m_{W,Z,h}\sim 100$ GeV so that a Little Hierarchy
(LH) appears with $\mu\ll m_{soft}$. 
We review twenty previously devised solutions to the SUSY $\mu$ problem 
and re-evaluate them in light of whether they are apt to support the LH. 
We organize the twenty solutions according to: 
1. solutions from supergravity/superstring constructions,
2. extended MSSM solutions,
3. solutions from an extra local $U(1)^\prime$ and 
4. solutions involving Peccei-Quinn (PQ) symmetry and axions.
Early solutions would invoke a global Peccei-Quinn symmetry
to forbid the $\mu$ term while relating the $\mu$ solution to solving the
strong CP problem via the axion. 
We discuss the gravity-safety issue pertaining to
global symmetries and the movement instead toward local gauge symmetries
or $R$-symmetries, either continuous or discrete. 
At present, discrete $R$-symmetries of order $M$ ($\mathbb{Z}_M^R$) 
which emerge as remnants of Lorentz symmetry
of compact dimensions seem favored. Even so, a wide variety of regenerative
mechanisms are possible, some of which relate to other issues such as the strong CP
problem or the generation of neutrino masses. 
We also discuss the issue of experimental verification or falsifiability of
various solutions to the $\mu$ problem.
Almost all solutions seem able to accommodate the LH.

\vspace*{0.8cm}

%\noindent PACS numbers: 12.60.Jv,14.80.Va,14.80.Ly

\end{abstract}

\end{titlepage}

\tableofcontents

\pagebreak

\section{Introduction: reformulating the $\mu$ problem for the LHC era}
\label{sec:intro}

Supersymmetry provides a solution to the Big Hierarchy problem-- 
why does the Higgs mass not blow up to the GUT/Planck scale--
via a neat cancellation of quadratic divergences which is required by
extending the Poincare group of spacetime symmetries to its maximal 
structure\cite{hier,wss}.
SUSY is also supported indirectly via the confrontation of data 
with virtual effects in that 
1. the measured gauge couplings unify under Minimal Supersymmetric Standard Model (MSSM) 
renormalization group evolution (RGE)~\cite{drw}, 
2. the measured value of $m_t$ falls in the range
required for a radiatively-driven breakdown of electroweak symmetry~\cite{ir},
3. the measured value of the Higgs boson mass falls squarely within the
narrow allowed range required by the MSSM~\cite{mhiggs,h125} and
4. the measured values of $m_W$ and $m_t$ favor the MSSM with heavy superpartners~\cite{Heinemeyer:2006px}.
In spite of these successes, so far no direct signal for SUSY has emerged at LHC 
leading to mass limits $m_{\tg}\agt 2$ TeV and $m_{\tst_1}\agt 1$ TeV while the rather large
value of $m_h\simeq 125$ GeV also seemingly requires multi-TeV highly mixed
top squarks~\cite{h125}. The new LHC Higgs mass measurement and sparticle 
mass limits seem to have exacerbated the so-called Little Hierarchy problem (LHP)~\cite{bs}:
why doesn't the Higgs mass blow up to the soft SUSY breaking scale
$m_{soft}\agt $several TeV, or what stabilizes the apparent hierarchy
$m_h\ll m_{soft}$? The LHP opens up the naturalness question:
how can it be that the weak scale $m_{weak}\sim m_{W,Z,h}\sim 100$ GeV without 
unnatural fine-tunings of dimensionful terms in the MSSM Lagrangian?

The most direct link between the magnitude of the weak scale and
the SUSY Lagrangian comes from minimization of the MSSM Higgs potential
to determine the Higgs field vevs~\cite{wss}. 
A straightforward calculation\cite{wss} reveals that
\be
m_Z^2/2= \frac{m_{H_d}^2+\Sigma_d^d-(m_{H_u}^2+\Sigma_u^u)\tan^2\beta}
{\tan^2\beta -1}-\mu^2\simeq -m_{H_u}^2-\Sigma_u^u(\tst_{1,2})-\mu^2
\label{eq:mzs}
\ee
where $\tan\beta\equiv v_u/v_d$ is the ratio of Higgs field vevs, 
$\mu$ is the SUSY conserving Higgs/higgsino mass term and $m_{H_{u,d}}^2$ 
are soft SUSY breaking up- and down-Higgs mass terms. 
The $\Sigma_u^u$ and $\Sigma_d^d$ terms contain a large assortment of loop corrections 
(see the Appendix of Ref.~\cite{rns2} for expressions) the largest of which
are usually the $\Sigma_u^u(\tst_{1,2})$ from the top-squark sector.

We can see immediately from the right-hand-side of Eq. \eqref{eq:mzs}
that if say one contribution is far larger than $m_Z^2/2$, then 
another (unrelated) term will have to be fine-tuned to compensate
so as to maintain $m_Z$ at its measured value. 
The {\it electroweak} fine-tuning measure $\Delta_{EW}$ has been introduced~\cite{rns2,rns1}--
\be
\Delta_{EW}\equiv max |largest\ term\ on\ RHS\ of\ Eq.~\eqref{eq:mzs}|/(m_Z^2/2)
\ee
-- to quantify the weak-scale fine-tuning required to maintain $m_Z$ at
its measured value. 
While a low value of $\Delta_{\rm EW}$ seems to be a necessary condition for
naturalness within the MSSM, the question is: is it also sufficient? It is argued 
in Ref's~\cite{dew,mt,seige,arno} that for {\it correlated} ({\it i.e.} inter-dependent)
soft terms as should occur in any more fundamental theory such as SUGRA with a 
well-specified SUSY breaking sector, or in string theory, then other measures
such as $\Delta_{\rm HS}\simeq \delta m_h^2/m_h^2$ and 
$\Delta_{\rm BG}\equiv max_i|\frac{\partial\log m_Z^2}{\partial\log p_i}|$ 
(where the $p_i$ are fundamental model parameters) 
collapse to $\Delta_{\rm EW}$ so that
$\Delta_{\rm EW}$ is sufficient as both an infra-red (IR) and ultra-violet (UV) 
fine-tuning measure. 
In contrast, theories with multiple independent soft parameters
may be susceptible to further fine-tunings which would otherwise cancel in a more 
fundamental theory. It should be recalled that in the multi-soft-parameter effective theories such
as CMSSM/mSUGRA, NUHM2 etc., the various soft parameters are introduced to parametrize one's 
ignorance of the SUSY breaking sector such that some choice of soft parameters will reflect
the true choice in nature. However, in no sense are the multi-soft-parameter theories
expected to be fundamental. Thus, in this paper we will adopt $\Delta_{\rm EW}$ as a 
measure of naturalness in fundamental theories with the MSSM as the weak scale effective theory.
In Ref.~\cite{upper}, it is shown that the fine-tuning already turns on 
for values of $\Delta_{\rm EW}\sim 20-30$. We will adopt a value of $\Delta_{\rm EW}<30$ as a
conservative choice for natural models of SUSY.

For a natural theory-- 
where $m_{W,Z,h}\sim 100$ GeV because the RHS contributions to Eq. \eqref{eq:mzs} 
are comparable to or less than the measured value of $m_Z^2/2$-- 
then evidently
\bi
\item $m_{H_u}^2(weak) \sim -(100-300)^2$ GeV$^2$ and
\item $|\mu |\sim 100-300$ GeV~\cite{ccn,bbh},
\item the largest of the radiative corrections (usually $\Sigma_u^u(\tst_{1,2})$) 
are not too large.
\ei

The first of these conditions pertains to the soft SUSY breaking sector. It can be achieved
for multi-TeV values of high-scale soft terms (as required by LHC limits) by 
radiatively driving $m_{H_u}^2$ from large, seemingly unnatural high scale values 
to a natural value at the weak scale. 
Thus, a high scale value of $m_{H_u}^2(\Lambda =m_{GUT} )$ must be selected such that
electroweak symmetry is barely broken. 
While this may seem to be a tuning in itself, 
such a selection seems to automatically emerge from SUSY within the 
string-landscape picture~\cite{DD,bbss}. 
In this scenario, there is a statistical draw towards large soft terms
which must be balanced by the anthropic requirement that EW symmetry be 
properly broken and with a weak scale magnitude not too far from its measured value\cite{don}.
The balance between these two tendencies pulls $m_{H_u}^2(m_{GUT} )$ 
to such large values that EW symmetry is barely broken.

The third of the above conditions-- 
that $\Sigma_u^u(\tst_{1,2})\sim 100-300$ GeV--
is achieved for third generation squark soft terms in the several TeV range 
along with a large trilinear soft term $A_t$ 
(as is expected in gravity-mediation models). 
These same conditions which reduce the $\Sigma_u^u(\tst_{1,2})$ values also 
increase the Higgs mass to its measured value $m_h\sim 125$ GeV~\cite{rns1,rns2}.

The second condition-- that the superpotential $\mu$ parameter
 is of order the weak scale-- brings up the famous SUSY $\mu$ problem~\cite{Polonsky:1999qd}:
since $W_{\rm MSSM}\ni\mu H_u H_d$ is SUSY preserving, naively one expects the 
dimensionful parameter $\mu$ to be of order $m_P\simeq 2.4\times 10^{18}$ GeV
while phenomenology requires $\mu\sim m_{weak}$. 
In this paper, we focus attention on the SUSY $\mu$ problem as occurs in 
gravity-mediation. The SUSY $\mu$ problem in gauge-mediated supersymmetry breaking (GMSB) is
summarized in Ref.~\cite{GR}. In GMSB, since the trilinear soft terms are 
expected to be tiny, then sparticle masses must become huge with
highly unnatural contributions to the weak scale in order to accommodate
a light Higgs boson with $m_h\simeq 125$ GeV~\cite{djouadi,bbm}.\footnote{
We also do not consider SUSY models with non-holonomic soft terms\cite{ross} or 
multiple $\mu$ terms; 
it is not clear whether such models have viable UV completions\cite{nelson,martin}.}

There are two parts to solving the SUSY $\mu$ problem:
\bi
\item First, one must forbid the appearance of $\mu$, usually via
some symmetry such as Peccei-Quinn (PQ) or better a 
continuous or discrete gauge or $R$-symmetry, and then 
\item re-generate $\mu$ at the much lower weak scale $|\mu |\sim 100-300$ GeV (the lower the more natural) 
via some mechanism such as symmetry breaking.
\ei
Many solutions to the SUSY $\mu$ problem have been proposed, 
and indeed in Sec. \ref{sec:rev} we will review twenty of these. 
In most of these solutions, the goal (for gravity-mediation) 
was to re-generate $\mu\sim m_{3/2}$ where $m_{3/2}$ is the gravitino mass which arises from
SUGRA breaking and which sets the mass scale for the soft SUSY breaking terms\cite{sugra}. 
When many of these $\mu$ solutions were proposed-- well before the LHC era-- 
it was commonly accepted that $m_{3/2}\sim m_{weak}$ which would also solve the SUSY 
naturalness problem. However, in light of the above discussion, the SUSY $\mu$ problem
needs a reformulation for the LHC era: any solution to the SUSY $\mu$ problem should first
forbid the appearance of $\mu$, but then re-generate it at the weak scale, 
{\it which is now hierarchically smaller than the soft breaking scale}:
\begin{equation}
|\mu |\sim m_{weak}\sim 100-300\ {\rm GeV}\ll m_{soft}\sim {\rm multi-TeV}\alt m_{3/2} .
\label{eq:muLHP}
\end{equation}

Our goal in this paper is to review various proposed solutions to the SUSY $\mu$ problem
and confront them with the Little Hierarchy as established by LHC data and as 
embodied by Eq. \ref{eq:muLHP}. While many solutions can be {\it tuned} to maintain the 
Little Hierarchy, others may offer compatibility with or even a mechanism to generate 
Eq. \ref{eq:muLHP}. Thus, present LHC data may be pointing to favored solutions to
the SUSY $\mu$ problem which may be reflective of the way nature actually works.

With this end in mind, in Sec. \ref{sec:rev} we will review a variety of 
mechanisms which have been offered as solutions to the SUSY $\mu$ problem. 
We organize the twenty solutions according to: 
\begin{itemize}
\item solutions from supergravity/superstring constructions,
\item extended MSSM solutions,
\item solutions from an extra local $U(1)^\prime$ and 
\item solutions involving Peccei-Quinn (PQ) symmetry and axions.
\end{itemize}
Many of these solutions tend to relate the $\mu$ parameter to the scale of soft SUSY breaking 
which would place the $\mu$ parameter
well above the weak scale and thus require significant EW fine-tuning. 
One such example is the original Kim-Nilles (KN)~\cite{kn} model (Subsec. \ref{ssec:kn}) 
which generates a $\mu$ parameter $\mu\sim v_{PQ}^2/m_P$ and 
relates $v_{PQ}\sim m_{hidden}$ (where $m_{hidden}$ is a mass scale associated with
hidden sector SUGRA breaking) and thus obtains  
$\mu\sim v_{PQ}^2/m_P \sim m_{hidden}^2/m_P \sim m_{3/2}$. 
However, the LHP can also be accomodated by allowing for $v_{PQ} \ll m_{hidden}$ 
so that $\mu \ll m_{3/2}$. While KN allows this possibility to be implemented
``by hand'', the later MSY~\cite{msy}, CCK~\cite{cck} and SPM~\cite{spm} models 
(Subsec. \ref{ssec:radpq}) implement radiative PQ breaking as a 
consequence of SUSY breaking  with the result that $v_{PQ} \ll m_{hidden}$ 
and hence $\mu\ll m_{soft}$~\cite{radpq}.

A prominent criticism of the $\mu$ solutions based on the existence of a global PQ
or discrete symmetry is that such symmetries are incompatible with 
gravity at high scales~\cite{nohair,wormhole,suss,dob,km_r}, 
{\it i.e.} that including the presence of gravity 
could spoil any global or discrete symmetries which may be postulated.
In Subsec. \ref{sssec:grav}, we discuss possible ways around the gravity 
spoliation of global or discrete symmetries. 
The MBGW model~\cite{bgw2} (Subsec. \ref{ssec:mbgw}) adopts a gravity-safe PQ symmetry 
thanks to a more fundamental discrete gauge symmetry $\mathbb{Z}_{22}$ 
and also generates PQ breaking from SUSY breaking, albeit not radiatively.

An attractive alternative to the discrete or continuous gauge symmetry 
resides in the possibility of a
discrete or continuous $R$ symmetry. Several discrete $R$-symmetries 
are possible which are anomaly-free (up to a Green-Schwarz term), 
forbid the $\mu$ parameter and other dangerous proton decay operators, and are 
consistent with an underlying grand unification structure\cite{lrrrssv1,lrrrssv2}.
Such discrete $R$-symmetries are expected to arise from compactification of
extra dimensions in string theory.
The $\mathbb{Z}_4^R$ symmetry stands out as a particularly simple approach that also 
leads to exact $R$-parity conservation. If one seeks to relate a gravity-safe 
PQ solution to the strong CP problem with a solution to the $\mu$ problem, 
then two hybrid models based on $\mathbb{Z}_{24}^R$ are examined (Subsec. \ref{ssec:hybrid}).
In this case, the PQ symmetry arises as an accidental approximate global
symmetry which emerges from the more fundamental discrete $R$ symmetry.
Here, the PQ breaking is generated through a large negative soft term 
and not radiatively.  

In Sec. \ref{sec:exp} we discuss the issue of experimental testability 
and distinguishability of various solutions to the $\mu$ problem. 
In Sec. \ref{sec:conclude}, we present a convenient Table \ref{tab:overview} 
which summarizes our review. 
Then we draw some final conclusions. 
Some pedogogical reviews providing an in-depth overview of 
supersymmetric models of particle physics can be found in Ref's \cite{wss}. 

\section{A review of some solutions to the SUSY $\mu$ problem}
\label{sec:rev}

In this Section, we review some solutions to the SUSY $\mu$ problem.
In the solutions reviewed here, the $\mu$-term is typically generated 
by breaking the symmetry which originally prohibits the $\mu$-term at the tree-level.
Depending on the source of such symmetry breaking,
we categorize the solutions according to 1. those from supergravity/superstring models, 
2. those from (visible-sector) extensions of the MSSM, 
3. those including an extra local $U(1)^\prime$ and 4. 
those which include also a solution to the strong CP problem 
with Peccei-Quinn symmetry breaking.

\subsection{Solutions in supergravity/string construction}

\subsubsection{Giudice-Masiero (GM)}
\label{ssec:gm}

In supergravity models the K\"ahler function $G=K+\log |W|^2$ is written
in terms of the real K\"ahler potential $K$ and the holomorphic 
superpotential $W$. 
If we posit some symmetry (PQ or $R$-symmetry are suggested in Ref. \cite{gm}) 
to forbid the usual MSSM $\mu$ term, then one may regenerate it via the
Higgs fields coupling to hidden sector fields $h_m$ via non-renormalizable 
terms in K~\cite{gm}:
\be
K\ni H_u^\dagger H_u+H_d^\dagger H_d +\left(\frac{\lambda_\mu}{m_P}H_u H_d h^\dagger +h.c.\right) .
\ee
If we arrange for SUSY breaking in the hidden sector, then the auxilliary 
component of $h$ develops a vev $\langle F_h\rangle\sim m_{hidden}^2$ 
so that the gravitino gets a mass $m_{3/2}\sim m_{hidden}^2/m_P$.
A $\mu$ term is generated of order
\be
\mu_{\rm eff}=\lambda_{\mu}\frac{\langle F_h^*\rangle}{m_P}\sim \lambda_{\mu} m_{hidden}^2/m_P\sim \lambda_{\mu} m_{3/2}\sim m_{soft} .
\ee
Thus, in the GM case, the $\mu$ parameter arises which is 
typically of order the soft breaking scale unless the coupling 
$\lambda_{\mu}$ is suppressed at the $\sim 0.01-0.1$ level.

\subsubsection{Casas-Munoz (CM)}
\label{ssec:cm}

Casas and Munoz~\cite{cm} propose a string theory inspired solution to the 
SUSY $\mu$ problem.
In string theory, dimensionful couplings such as $\mu$ are already forbidden 
by the scale invariance of the theory so no new symmetries are needed to forbid it.
They begin with a superpotential of the form
\be
W=W_0+\lambda_{\mu}W_0H_uH_d/m_P^2
\label{eq:cm_W}
\ee
where $W_0$ is the usual superpotential of the MSSM (but without the $\mu$ term) along with the
hidden sector component which is responsible for SUSY breaking: $W_0=W_0^{vis}(z_i)+W_0^{hid}(h_m)$
where the $z_i$ comprise visible sector fields while the $h_m$ denote hidden sector fields.
While the scale-variant $\mu$ term is forbidden in $W_0^{vis}$, the non-renormalizable
contribution in Eq. \eqref{eq:cm_W} is certainly allowed and, 
absent any symmetries which could forbid it, probably mandatory.
Under, for instance, $F$-term SUSY breaking in the hidden sector, then $W_0^{hid}$ gains a vev
$\langle W_0^{hid}\rangle\sim m_{hidden}^2m_P$ (as is easy to see in the simplest Polonyi model 
for SUSY breaking with $W_{Polonyi}=m_{hidden}^2(h+\beta m_P)$ where $\beta$ is a dimensionless
constant). Under these conditions, then a $\mu$ term develops with
\be
\mu_{\rm eff}\sim \lambda_{\mu}m_{hidden}^2/m_P\sim \lambda_{\mu} m_{3/2} \sim m_{soft}.
\label{eq:cm_mu}
\ee
Ref.~\cite{cm} goes on to show that the CM solution can easily emerge in models 
of SUSY breaking due to hidden sector gaugino condensation at some intermediate mass scale 
$\Lambda_h$ (where then we would associate $m_{hidden}^2\simeq \Lambda_h^3/m_P$).

A benefit of the CM solution is that it should be consistent with any stringy UV completion~\cite{saul}
as it avoids the presence of some global (PQ) symmetry. A possible drawback to CM is that the
$\mu$ term is naturally expected to be of order $m_{soft}$ instead of $m_{weak}$ 
unless $\lambda_{\mu}$ is suppressed (as in GM).
One way to falsify the CM solution would be to discover a DFSZ-like axion with
consistent mass and coupling values. Such a discovery would exclude the second term in
Eq. \eqref{eq:cm_W} since it would violate the PQ symmetry.

\subsubsection{$\mu$ and a big hierarchy from approximate $R$-symmetry}
\label{ssec:Rsym}

In string theory models, approximate $R$-symmetries are expected to
develop from overall Lorentz symmetry of the 10-dimensional spacetime 
when compactified to four dimensions. Under a continuous $U(1)_R$
symmetry, the superspace co-ordinates transform non-trivially and hence 
so do the bosonic and fermionic components of superfields. 
Thus, these symmetries can be linked to overall Lorentz symmetry 
where also bosons and fermions transform differently. 

Under exact $R$-symmetry and supersymmetry, then the superpotential
$\mu$ term is forbidden since the gauge-invariant bilinear term of Higgs pair $H_uH_d$
carries zero $R$-charge while the superpotential must have $R_W=+2$.
However, $H_uH_d$ may couple to various other superfields
$\phi_i$ which carry non-trivial $R$-charges so that 
\be
W\ni P_\mu(\phi_i )H_uH_d 
\ee
where $P_\mu (\phi_i )$ is a sum over monomials in  the fields $\phi_i^n$.
Unbroken $R$-symmetry requires a vanishing $\langle P_\mu (\phi_i )\rangle$
but if the $R$-symmetry is approximate then non-vanishing $P_\mu (\phi_i )$
contributions will develop at higher orders in powers of the field vevs
$\langle (\phi_i/m_P)\rangle \alt 1$. Thus, a mild hierarchy in the
field vevs $\langle \phi_i/m_P\rangle \alt 1$, when raised to higher powers 
$\langle (\phi_i/m_P)^{n_i}\rangle \ll 1$, can 
generate a much larger hierarchy of scales~\cite{kappl}. 
In this solution to the $\mu$ problem, which is essentially a UV completion of the
CM solution, then $\mu\sim m_{3/2}\sim \langle W\rangle$ is expected to arise.

\subsubsection{Solution via the discrete $R$-symmetry $\mathbb {Z}_4^R$}
\label{ssec:Z4R}

A particularly attractive way to solve the $\mu$ problem in some string
constructions is via a discrete Abelian $R$-symmetry $\mathbb{Z}_4^R$~\cite{cckR,hnp,DineR}. 
Such $R$-symmetries may arise as discrete remnants of the Lorentz
symmetry of extra dimensional ($d=10$) models upon compactification to $d=4$.
In Ref.~\cite{babu}, the $\mathbb{Z}_4^R$ symmetry was invoked to forbid the 
$\mu$ term
as well as dimension-4 baryon- and lepton-number violating operators
while dangerous dimension-5 operators leading to proton decay 
are highly suppressed~\cite{lrrrssv1,lrrrssv2}.
The desirable Weinberg neutrino mass operator is allowed. 
The $\mathbb{Z}_4^R$ charges
are assigned so that all anomalies cancel by including Green-Schwarz terms
(and extra $R$-charged singlets for gravitational anomalies). 
The $R$-charge assignments for the discrete $R$-symmetry $\mathbb{Z}_4^R$ 
are shown in the second row of Table \ref{tab:Z4R}.
\begin{table}[!htb]
\renewcommand{\arraystretch}{1.2}
\begin{center}
\begin{tabular}{c|cccccccc}
multiplet & $H_u$ & $H_d$ & $Q_i$ & $L_i$ & $U_i^c$ & $D_i^c$ & $E_i^c$ & $N_i^c$ \\
\hline
 $\mathbb{Z}_{4}^R$ charge & 0 & 0  & 1 & 1 & 1 & 1 & 1 & 1 \\
\hline
\end{tabular}
\caption{ $\mathbb{Z}_{4}^R$ charge assignments for various superfields of the 
LRRRSSV model\cite{lrrrssv1}.
}
\label{tab:Z4R}
\end{center}
\end{table} 

The charge assignments are consistent with embedding the matter superfields 
into a single ${\bf 16}$ of $SO(10)$ while the split Higgs multiplets would 
arise from Wilson-line breaking of gauge symmetry. 
The $\mathbb{Z}_4^R$ symmetry may be broken via non-perturbative effects such as gaugino 
condensation breaking of SUGRA in the hidden sector so that 
a gravitino mass $m_{3/2}$ is induced along with soft terms $m_{soft}\sim m_{3/2}$. 
A $\mu$ term may arise via GM (Sec. \ref{ssec:gm}) and/or CM (Sec.~\ref{ssec:cm}) so that
$\mu\sim\langle W\rangle/m_P^2\sim m_{3/2}\sim m_{soft}$. 
Although the discrete $\mathbb{Z}_4^R$ $R$-symmetry is broken, the discrete
matter/$R$-parity remains unbroken so that the LSP remains absolutely stable.
This sort of solution to the $\mu$ problem is expected to be common in 
heterotic string models compactified on an orbifold~\cite{lrrrssv2}.
Other possibilities for $\mathbb{Z}_N^R$ with $N>4$ also occur\cite{lrrrssv2} 
and in fact any $N$ value is possible under anomaly cancellations provided one includes
additional exotic matter into the visible sector~\cite{Harigaya:2013vja}.

A further concern is that a spontaneously broken discrete symmetry 
may lead to formation of domain walls in the early universe which 
could dominate the present energy density of the universe~\cite{sikivie,Larsson:1996sp,dine}.
For the case of gravity mediation, the domain walls would be expected to 
form around the SUSY breaking scale $T\sim 10^{12}$ GeV. However, 
if inflation persists to lower temperatures, then the domain walls may be 
inflated away. It is key to observe that many mechanisms of baryogenesis
are consistent with inflation persisting down to temperatures of
$T\sim 10^6$ GeV~\cite{baryo}.

%\subsection{Solutions from string theory}

\subsubsection{String instanton solution}
\label{ssec:instanton}

In string theory models, it is possible for superpotential terms to arise 
from non-perturbative instanton effects. 
These are particularly well suited for
open strings in braneworld scenarios such as IIA and IIB string theory.
Intriguing applications of stringy instanton effects include the generation
of Majorana neutrino mass terms, generation of Yukawa couplings and 
generation of the $\mu$ term in the superpotential~\cite{ibanez_uranga,gw}.
In some D-brane models which include the MSSM at low energy, then the
superpotential $\mu$ term may be forbidden by $U(1)$ symmetries
but then it is generated non-perturbatively via non-gauge 
$D$-brane instanton effects.
In this case, then a $\mu$ term of the form
\be
W\sim \exp (-S_{\rm cl})M_s H_u H_d 
\ee
can be induced where then $\mu\simeq \exp (-S_{\rm cl})M_s$ and $M_s$ is the 
string mass scale. 
The exponential suppression leads to the possibility of a $\mu$ term 
far below the string scale. Of course, in this case one might expect 
the $\mu$ term to arise at any arbitrary mass scale below 
the string scale rather than fortuitously at the weak scale. 
If the $\mu$ term does arise at the weak scale from stringy instanton 
effects, then that value may act as an attractor such that 
soft terms like $m_{H_u}^2$ are pulled statistically to large values by the
string theory landscape, but not so large that EW symmetry doesn't break.
Then the weak scale value of $m_{H_u}^2$ is of comparable (negative) magnitude 
to $\mu$ (the naturalness condition) to ensure a universe with
anthropically required electroweak symmetry breaking~\cite{bbss}. 

\subsubsection{Mu solution in ${\rm G_2MSSM}$}
\label{ssec:g2mssm}

In Ref.~\cite{kane} (Acharya {\it et al.}), the authors consider 11-dimensional $M$-theory
compactified on a manifold of $G_2$ holonomy, and derive various 
phenomenological implications. 
They consider fields living in multiplets of 
$SU(5)$ so the doublet-triplet splitting problem is present.
As opposed to string theory models compactified on orbifolds, in $M$-theory
the matter fields live only in four dimensions so a different solution to the
$\mu$ problem is required. Witten suggested the existence of an additional 
discrete symmetry which forbids the $\mu$ term from appearing but which allows 
the Higgs triplets to gain large enough masses so as to evade proton decay
constraints~\cite{witten}. In Ref.~\cite{kane_mu}, it is shown that a $\mathbb{Z}_4$ symmetry is
sufficient to forbid the $\mu$ term and other dangerous RPV operators while
allowing massive Higgs triplets. The $\mathbb{Z}_4$ discrete symmetry is assumed to 
be broken via moduli stabilization so that a small $\mu$ term develops.

In the $G_2MSSM$, the gravitino gains mass from non-perturbative effects
(such as gaugino condensation) in the hidden sector so that 
$m_{3/2}\sim \Lambda_h^3/m_P^2\sim 10-200$ TeV. 
Matter scalar soft masses are expected at $m_\phi\sim m_{3/2}$ 
so should be very heavy 
(likely unnatural in the context of Eq. \eqref{eq:mzs}). 
In contrast, gauginos gain mass from the 
gauge kinetic function which depends on the vevs of moduli fields 
so they are expected to be much lighter: $m_{\lambda}\sim $TeV scale 
and in fact these may have dominant AMSB
contributions~\cite{amsb} (with comparable moduli-mediated SUSY breaking contributions) 
so that the wino may be the lightest of the gauginos. The dominant
contribution to the $\mu$ parameter arises from K\"ahler contributions
ala Giudice-Masiero and these are expected to be 
$\mu\sim c\frac{\langle S_i\rangle}{m_p}m_{3/2}\sim 0.1 m_{3/2}$ 
(where $c$ is some constant $\sim 1$) 
and thus is suppressed compared to scalar soft masses, 
but perhaps comparable to gaugino masses.

\subsection{Extended MSSM-type solutions}

\subsubsection{NMSSM: Added singlet with $\mathbb{Z}_3$ discrete symmetry}
\label{ssec:nmssm}

The case of adding an additional visible-sector gauge singlet superfield $S$ 
to the MSSM leads to the next-to-minimal SSM or NMSSM~\cite{nmssm}. 
Some motivation for the NMSSM can originate in string
theory models such as heterotic orbifolds where the $\mu$-term arises as 
an effective term from couplings of the Higgs pair to a singlet field~\cite{saul}.
Without imposing any symmetry to forbid singlet couplings,
we can write a generic NMSSM superpotential as follows:
\be
W_{NMSSM}=W_{MSSM}(\mu =0 )+\lambda_\mu S H_u H_d+
\xi_F S +\frac{1}{2}\mu_SS^2 + \frac{1}{3}\kappa S^3
\label{eq:gen_nmssm}
\ee
and corresponding soft terms
\be
{\cal L}_{soft}^{NMSSM}={\cal L}_{soft}^{MSSM}-(a_\lambda S H_uH_d+B\mu H_uH_d
+\frac{1}{3}a_\kappa S^3 +\frac{1}{2}b_SS^2+t S+c.c.)-m_S^2|S|^2 .
\label{eq:NMSSMsoft}
\ee
Here $W_{MSSM}(\mu=0)$ denotes the superpotential for the MSSM but 
without the $\mu$-term.
The tadpole $t$ in Eq. \eqref{eq:NMSSMsoft} may have destabilizing quadratic 
divergences and must be suppressed~\cite{bagger_etc}. 
A $\mathbb{Z}_3$ discrete symmetry is usually imposed 
wherein chiral superfields transform as $\phi\to e^{2\pi i/3}\phi$ which 
sends the dimensionful couplings $\xi_F$, $\mu$, $\mu_S$, $B\mu$, $b_S$ and $t$ to zero 
(only cubic couplings are allowed) at the 
expense of possibly introducing domain walls into the early universe 
after the electroweak  phase transition~\cite{nmssm_domain}.
(Some means of avoidance of domain walls are proposed in 
Ref's~\cite{Abel:1996cr}.) 
By minimizing the scalar potential, now including
the new singlet scalar $S$, then vevs $v_u$, $v_d$ and $v_s$ are induced.
An effective $\mu$ term emerges with
\be
\mu_{\rm eff}=\lambda_\mu v_s .
\ee
   
An attractive alternative choice for $\mu$-forbidding symmetry 
than the (perhaps ad-hoc) $\mathbb{Z}_3$ would be one of the 
anomaly-free discrete $R$-symmetries $\mathbb{Z}_4^R$ 
or $\mathbb{Z}_8^R$~\cite{lrrrssv2}. 
Like the $\mathbb{Z}_3$ discrete symmetry, the $\mathbb{Z}_8^R$ symmetry also forbids 
the dangerous divergent tadpole term. 
The  $\mathbb{Z}_4^R$ symmetry would allow the linear singlet term, but
it can be argued that in the effective theory the linear term appears 
when the fields with which the singlet field is coupled acquire VEVs. 
If these fields belong to the hidden sector, then the coupling will be 
suppressed by some high mass scale ranging as high as $m_P$ in the 
case of gravity-mediation. 
In this case the linear singlet term will be present but it will be 
highly suppressed~\cite{lrrrssv2}.
 
Thus, all the advantages of the $\mathbb{Z}_3$ discrete symmetry can be 
obtained by imposing instead either a $\mathbb{Z}_4^R$ or $\mathbb{Z}_8^R$ 
symmetry: this then  avoids the disadvantages--ad-hocness and introduction 
of domain walls into the early universe after electroweak phase transition-- 
inherent in the $\mathbb{Z}_3$ discrete symmetry.

The added singlet superfield $S$  in the NMSSM leads to new scalar and 
pseudoscalar Higgs fields 
which can mix with the usual MSSM Higgses for $v_s\sim v_{u,d}$. So far, 
LHC Higgs coupling measurements favor a SM-like Higgs so one might expect
$v_s\gg v_{u,d}$ which may lead one to an unnatural value of $\mu_{\rm eff}$.
The superfield $S$ also contains a spin-$1\over 2$ singlino $\tilde{s}$ 
which may mix with the usual neutralinos and might even be the 
LSP~\cite{balazs}. 
In the NMSSM, an additional Higgs quartic potential term is generated from the $F$-term of the singlet superfield, 
and thus the SM-like Higgs mass 125~GeV is explained more easily
without introducing large one-loop corrections.
This feature can make the NMSSM more attractive to those who are 
uncomfortable with an MSSM Higgs of mass $m_h\simeq 125$ GeV\cite{Hall:2011aa}.

%\cite{Feldstein:2004xi}

\subsubsection{nMSSM}
\label{sssec:nMSSM}

An alternative singlet extension of the MSSM is the Nearly-Minimal Supersymmetric Standard 
Model (nMSSM) (also sometimes called Minimal Nonminimal Supersymmetric Standard Model or 
MNSSM)~\cite{Panagiotakopoulos:1999ah,Panagiotakopoulos:2000wp}.
The nMSSM, like the NMSSM, solves the $\mu$ problem via an added singlet superfield $S$. 
But in the nMSSM, the model is founded on a discrete $R$-symmetry either $\mathbb{Z}_5^R$ or 
$\mathbb{Z}_7^R$. Discrete $R$-charge assignments for $\mathbb{Z}_5^R$ are shown in Table \ref{tab:nMSSM}.
The tree level superpotential is given by
\be
W_{nMSSM}\ni \lambda_\mu SH_uH_d+f_uQH_uU^c+f_dQH_d D^c+f_{\ell}LH_dE^c+f_{\nu}LH_uN^c
+\frac{1}{2} M_N N^cN^c \nonumber \\
\ee
so that unlike the NMSSM with $\mathbb{Z}_3$ symmetry, the $\kappa S^3$ term is now forbidden. 
This is why the model is touted as a more minimal extension of the MSSM.
The discrete $R$ symmetry is broken by SUSY breaking effects in gravity-mediation.
Then, in addition to the above terms, an effective 
potential tadpole contribution
\be
W_{nMSSM}^{tad}\ni \xi_F S 
%\sim \left(\frac{1}{16\pi^2}\right)^nM_pM_{SUSY}^2S
\ee
is induced at six-loop or higher level where $\xi_F\sim m_{3/2}^2$  
(along with a corresponding soft SUSY breaking term).
Due to lack of the discrete global $\mathbb{Z}_3$ symmetry, the nMSSM then avoids the domain wall
and weak scale axion problems that might afflict the NMSSM.
\begin{table}[!htb]
\renewcommand{\arraystretch}{1.2}
\begin{center}
\begin{tabular}{c|ccccccccc}
multiplet & $H_u$ & $H_d$ & $Q_i$ & $U_i^c$ & $D_i^c$ & $L_i$ & $E_i^c$ & $N^c$ & $S$ \\
\hline
$\mathbb{Z}_5^R$ & 2 & 2  & 4 & 6 & 6 & 4 & 6 & 6 & 3 \\
\hline
\end{tabular}
\caption{Charge assignments for various superfields of nMSSM with a 
$\mathbb{Z}_5^R$ discrete $R$-symmetry.
}
\label{tab:nMSSM}
\end{center}
\end{table}

Like the NMSSM, the nMSSM will include added scalar and pseudoscalar Higgs particles along with
a fifth neutralino. However, due to lack of the $S$ self-coupling term and presence of the tadpole term,
the mass eigenstates and couplings of the added matter states will differ from the 
NMSSM~\cite{Dedes:2000jp,Menon:2004wv,Barger:2006dh,Barger:2006kt,Cao:2009ad}.
The neutralino in the nMSSM is very light, mostly below 50 GeV,
but it is hard to get lower than 30 GeV due to the dark matter relic density constraint. 
Since the neutralinos are so light it is very 
likely that a chargino will decay into either a MSSM-like $\chi_2^0$ or a singlino $\chi_1^0$, giving rise 
to a 5 lepton final state. 
A further decay of the neutralino can give rise to a 7 lepton state. 
These kinds of multilepton events are more likely in the 
nMSSM than in the NMSSM. 
Also, since in the nMSSM the neutralino can be so light, 
then deviations in Higgs boson $h$ decay branching fractions 
become more likely than in the case of the 
NMSSM\cite{Barger:2006dh,Barger:2006kt}.

\subsubsection{Mu-from-nu SSM ($\mu\nu$SSM)}
\label{ssec:munuSSM}

The $\mu$-from-$\nu$SSM ($\mu\nu$SSM)~\cite{LopezFogliani:2005yw} 
is in a sense a more minimal version of the NMSSM in that it makes use of the 
gauge singlet right-hand-neutrino 
superfields $N^c_i$ to generate a $\mu$ term.
The $\mu\nu$SSM first requires a $\mathbb{Z}_3$ symmmetry to forbid the usual 
$\mu$ term (and also a usual Majorana neutrino mass term $M_iN^cN^c$). 
The superpotential is given by
\bea
W &\ni &f_uQH_uU^c+f_dQH_d D^c+f_{\ell}LH_dE^c+f_{\nu}LH_uN^c \nonumber \\
& +& \lambda_{\mu i} N^c_i H_u H_d+{1\over 3}\kappa_{ijk}N_i^cN_j^cN_k^c .
\eea
If the scalar component of one of the RHN superfields 
$\tilde{\nu}_{Ri}$ of $N_i^c$ gains a weak scale vev,
then an effective $\mu$ term develops:
\bea
\mu_{\rm eff} = \lambda_{\mu i} \langle\tilde{\nu}_{Ri}\rangle
\eea
along with a weak scale Majorana neutrino mass term 
$M_{Njk}\sim\kappa_{ijk}\langle\tilde{\nu}_{Ri}\rangle$.
By taking small enough neutrino Yukawa couplings, then a weak scale
see-saw develops which can accommodate the measured neutrino
masses and mixings.

The $\mu\nu$SSM develops bilinear $R$-party violating terms via the
superpotential $f_\nu LH_uN^c$ term so that the lightest $\mu\nu$SSM particle
is not stable and doesn't comprise dark matter: $\tilde{\chi}_1^0\to W^{(*)}\ell$
and other modes. 
As an alternative, a gravitino LSP is suggested with age longer than the 
age of the universe:
it could decay as $\tilde{G}\to\nu\gamma$ and possibly yield gamma ray signals
from the sky~\cite{Choi:2009ng}. 
The phenomenology of the $\mu\nu$SSM also becomes more
complex: now the neutrinos inhabit the same mass matrix as neutralinos,
leptons join charginos in another mass matrix and Higgs scalars and sneutrinos
inhabit a third mass matrix 
(albeit with typically small mixing effects).
Collider signals are strongly modified from usual MSSM 
expectations~\cite{Fidalgo:2011ky}.

While the $\mu\nu$SSM may be considered the most minimal model to solve 
the $\mu$ problem, it suffers the same $\mathbb{Z}_3$ domain wall problem as the NMSSM
(and perhaps the same routes to avoidance~\cite{Abel:1996cr}). 
Also, in the context of GUTs, the role that the $N_i^c$ field plays in
the {\bf 16}-dimensional spinor of $SO(10)$ woud have to be abandoned.

\subsection{$\mu$ from an extra local $U(1)^\prime$}
\label{ssec:u1}

In this class of models~\cite{cvetic,xt,mw,cp,arvanitaki}, 
a SM singlet superfield $S$ is introduced which is charged under a new 
$U(1)^\prime$ gauge interaction, so
terms with mass dimensions in Eq.~\eqref{eq:gen_nmssm} are forbidden.
% by 
%a new gauge $U(1)^\prime$ symmetry instead of discrete symmetries shown in previous subsections.
%as in the NMSSM, 
%one adds an additional singlet field $S$ (along with additional exotica) to the MSSM. 
%But this time, instead of a global symmetry forbidding 
%the $\mu$ term, an extra local $U(1)^\prime$ gauge symmetry is postulated.
Due to the $U(1)^\prime$ gauge charges of $S$, 
the cubic coupling $S^3$ is also absent.
We will see below three representative realizations of this class of model.

\subsubsection{CDEEL model}
\label{sssec:CDEEL}

Cvetic-Demir-Espinosa-Everett-Langacker~\cite{cvetic} (CDEEL) 
propose a $U(1)^\prime$ extended gauge symmetry model as emblematic of 
fermionic orbifold string compactifications. 
While the usual $\mu$ term is forbidden by the extended gauge symmetry, 
the superpotential term
\be
W\ni\lambda_{\mu} S H_u H_d
\ee
is allowed and under $U(1)^\prime$ breaking then $S$ develops a vev $\langle S\rangle\sim m_{weak}$ 
such that a $\mu$ term is generated $\mu_{\rm eff}=\lambda_{\mu}\langle S\rangle$
along with an additional weak scale $Z^\prime$ gauge boson.
Forbidding the $\mu$ term via a gauge symmetry avoids the gravity spoliation/global symmetry problem. 
In addition, the $\mu$ term is linked to EW
symmetry breaking and this would be expected to occur at $m_{weak}$ rather than $m_{soft}$. 
The $U(1)^\prime$ breaking can occur either via
large soft SUSY breaking trilinear couplings or via radiative corrections driving
certain mass-squared terms negative.
A way to test this class of models, in the exotica decoupling limit,
is to search for new $Z^\prime$ gauge bosons with exotic decays to
light higgsinos~\cite{cp}.

To maintain anomaly cancellation, a variety of (intermediate scale) exotic quark and lepton fields
must be introduced along with extra SM gauge singlets.
If these new states come in GUT representations,
then gauge coupling unification can be maintained.
A set of possible $U(1)^\prime$ gauge charges are listed in Table~\ref{tab:cp}.
\begin{table}[!htb]
\renewcommand{\arraystretch}{1.2}
\begin{center}
\begin{tabular}{c|cccccccc}
multiplet & $H_u$ & $H_d$ & $Q_i$ & $U_i^c$ & $D_i^c$ & $L_i$ & $E_i^c$ & $S$ \\
\hline
$(2\sqrt{10})Q^\prime$ & -2 & -3  & 1 & 1 & 2 & 2 & 1 & 5 \\
\hline
\end{tabular}
\caption{Charge assignments for various superfields of a $U(1)^\prime$ model~\cite{cp,mw}.
}
\label{tab:cp}
\end{center}
\end{table}
\\

\subsubsection{sMSSM model}
\label{sssec:sMSSM}

An alternative $U(1)^\prime$-extended MSSM (abbreviated as sMSSM)\cite{Erler:2002pr,Han:2004yd} 
also solves the $\mu$ problem by invoking multiple SM singlet superfields charged under $U(1)^\prime$ symmetry.
%an additional $U(1)^\prime$ gauge symmetry 
In this model, a visible-sector singlet field $S$ directly couples to Higgs doublets
but avoids stringent constraints on having an 
additional weak scale $Z^\prime$ gauge boson by introducing as well a 
{\it secluded sector} containing three additional singlets $S_1,\ S_2,\ S_3$ charged under $U(1)^\prime$.
The superpotential is given by
\be
W_{sMSSM}\ni\lambda_{\mu} S H_u H_d+\lambda_s S_1S_2S_3
\ee
so that the secluded sector has a nearly $F$- and $D$-flat scalar potential. 
The $U(1)^\prime$ and electroweak symmetry breaking then occurs as a result of SUSY breaking
$A$-terms. 
Then the secluded sector scalars can obtain vevs much larger than the weak scale; if also the trilinear singlet coupling $\lambda_s$ is small, 
then the additional $Z^\prime$ essentially decouples. 
Nonetheless, additional Higgs and singlinos appear in the weak scale effective theory 
so that this model phenomenologically resembles the nMSSM 
(described in Subsec.~\ref{sssec:nMSSM}) which has very different
manifestations from what is expected from the CDEEL $U(1)^\prime$ model.

\subsubsection{HPT model}
\label{sssec:HPT}

The Hundi-Pakvasa-Tata (HPT) model~\cite{xt} also solves the SUSY $\mu$ problem
by positing an additional $U(1)^\prime$ gauge symmetry in a supergravity 
context. 
The $U(1)^\prime$ charges of the multiplets in the HPT scheme are shown
in Table \ref{tab:hpt}. 
With these $U(1)^\prime$ charge assignments, the $\mu$ term
is forbidden in the superpotential but (unlike the CDEEL model) a dim-4 term 
as $\mu$ solution \`a la Kim-Nilles is allowed:
\be
W\ni\lambda_{\mu} S^2 H_u H_d/M_p .
\ee
The $U(1)^\prime$ gauge symmetry also forbids trilinear RPV couplings
and dangerous $p$-decay operators.
When the $U(1)^\prime$ breaks (at an intermediate scale $Q\sim 10^{11}$ GeV), 
the $S$ field acquires a vev to yield an effective $\mu$ parameter of the 
required magnitude.

A distinctive feature of the HPT model is that a bilinear RPV (bRPV) term, $LH_u$ is allowed at
the right magnitude so as to generate phenomenologically-allowed neutrino masses~\cite{valle}.
The desired pattern of neutrino masses and mixing angles
are also accommodated through radiative corrections.
The bRPV leads to an unstable lightest neutralino which decays via 
$\tilde{\chi}_1^0\to \ell W^{(*)}$ or $\nu Z^{(*)}$ and may lead to displaced vertices
in collider events. Dark matter must be comprised of some other particles ({\it e.g.} axions).
Also, the $U(1)^\prime$ is broken at the intermediate scale $Q\sim 10^{11}$ GeV so that
the additional $Z^\prime$ has a mass far beyond any collider reach.

Since solving the $\mu$ problem as well as generating the neutrino mass scale of suitable order
requires introduction of a new gauge group $U(1)^\prime$, care must be taken so that
associated anomalies are cancelled. 
Anomaly cancellation requires introducing various additional exotic fields
including color triplets $K_i$ and $K_i^\prime$ states. The lightest of these 
leads to stable weak-scale exotic hadrons which may also yield 
highly-ionizing tracks at collider experiments. 
In the HPT scheme, gauge coupling unification may be upset. 
\begin{table}[!htb]
\renewcommand{\arraystretch}{1.2}
\begin{center}
\begin{tabular}{c|cccccccc}
multiplet & $H_u$ & $H_d$ & $Q_i$ & $U_i^c$ & $D_i^c$ & $L_i$ & $E_i^c$ & $S$ \\
\hline
$Q^\prime$ & 25 & -31  & 0 & -25 & 31 & 2 & 29 & 3 \\
\hline
\end{tabular}
\caption{Charge assignments for various superfields of the HPT $U(1)^\prime$ 
supergravity model~\cite{xt}.
}
\label{tab:hpt}
\end{center}
\end{table}

\subsection{Solutions related to Peccei-Quinn symmetry breaking}

In this Subsection, we examine natural $\mu$-term solutions related to 
the PQ symmetry used to solve the strong CP problem.
In this class of models, the $\mu$-term is forbidden by the PQ symmetry, 
but generated once the PQ symmetry is spontaneously broken.
Then the model also provides a solution to the strong CP problem and 
generates axion dark matter.
In Subsec.~\ref{ssec:kn}, \ref{ssec:ckn}, and \ref{ssec:king},
we review $\mu$-term generation models with various sources of PQ breaking.

Meanwhile, imposing a global symmetry causes the `quality' 
issues of the symmetry which may spoil the PQ solution to the strong CP problem,
since global symetries are not protected from quantum gravity effects.
In Subsec.~\ref{sssec:grav}, we discuss a criterion for protecting the PQ solution to the strong CP problem,
and in Subsec.~\ref{sssec:gra_rsym} we present examples based on 
discrete $R$-symmetries which satisfy the gravity-safety criterion
and can be considered as generating an accidental, approximate PQ symmetry.
Also, we review the natural Higgs-flavor-democracy (HFD) solution 
which contains an approximate PQ symmetry
from a discrete symmetry in Subsec.~\ref{sssec:hfd}.

Finally, we review $\mu$-term generation by breaking of PQ symmetry from SUSY breaking:
radiative breaking of PQ symmetry (Subsec.~\ref{ssec:radpq}), breaking of an accidental approximate PQ symmetry from a gauged $U(1)_R$ symmetry (Subsec.~\ref{ssec:CCL}) and a $\mathbb{Z}_{22}$ discrete gauge symmetry (Subsec.~\ref{ssec:mbgw}) by a large negative trilinear term.

\subsubsection{Kim-Nilles solution}
\label{ssec:kn}

Kim and Nilles (KN)~\cite{kn} presented the first formulation 
of the SUSY $\mu$ problem along with a proposed solution. 
In KN, it is proposed that there exists a global Peccei-Quinn (PQ) symmetry $U(1)_{PQ}$ 
which is needed at first as a solution to the strong CP problem. 
The PQ symmetry is implemented in the context of the supersymmetrized version
of the DFSZ~\cite{dfsz} axion model\footnote{In the DFSZ axion model~\cite{dfsz}, the SM is extended to include two Higgs doublets which then couple to singlets
which contain the axion.} wherein the Higgs multiplets carry PQ
charges {\it e.g.} $Q_{PQ}(H_u)=Q_{PQ}(H_d)=-1$ so that the $\mu$ term is 
forbidden by the global $U(1)_{PQ}$. 
Next, the Higgs multiplets are coupled 
via a non-renormalizable interaction to a SM gauge singlet field $X$
which carries a PQ charge $Q_{PQ}(X)=+2/(n+1)$:
\begin{equation}
W_{\mu}\ni \frac{\lambda_{\mu}}{m_P^n}X^{n+1}H_uH_d 
\end{equation}
for $n\ge 1$. 

It is arranged to spontaneously break PQ by giving the $X$ field a vev $\langle X\rangle$
which also generates a (nearly) massless axion $a$ which solves the strong CP problem.
To obtain cosmologically viable axions-- with 
$\langle X\rangle \sim 10^{11}$ GeV and with $m_p \simeq 2.4\times 10^{18}$ GeV, we can obtain 
the $\mu$ parameter of the order of $m_{3/2}$ only if $n = 1$ (for which $Q_{PQ}(X) = +1$).
The matter superfields also carry appropriate PQ charge so as to 
allow the MSSM trilinear superpotential terms: see Table \ref{tab:kn}.

\begin{table}[!htb]
\renewcommand{\arraystretch}{1.2}
\begin{center}
\begin{tabular}{c|cccccccccc}
multiplet & $H_u$ & $H_d$ & $Q_i$ & $L_i$ & $U_i^c$ & $D_i^c$ & $E_i^c$ & X & Y & Z \\
\hline
PQ charge & $-1$ & $-1$  & $+1$ & $+1$ & 0 & 0 & 0 & +1 & -1 & 0\\
\hline
\end{tabular}
\caption{PQ charge assignments for various superfields of the KN model
with $n=1$.
One may add multiples of weak hypercharge or $B-L$ to these so their values are not unique.
}
\label{tab:kn}
\end{center}
\end{table}

The intermediate PQ breaking scale can be gained from a PQ superpotential of the form:
\be
W_{PQ}=\lambda_{PQ}Z\left( XY-v_{PQ}^2\right).
\label{eq:knW}
\ee
The scalar components of $X$ and $Y$ develop vevs 
$\langle X\rangle =\langle Y\rangle = v_{PQ}$
such that a $\mu$ term is generated:
\be
\mu = \lambda_\mu \langle X\rangle^2/m_P .
\ee
This value of the $\mu$ term $\mu\sim \lambda_{\mu}v_{PQ}^2/m_P$ 
is to be compared to the soft breaking scale
in models of gravity-mediation: $m_{soft}\sim m_{3/2}\sim m_{hidden}^2/m_P$.
Here, $v_{PQ}$ is identified as $v_{PQ} \sim m_{hidden}$ and thus $\mu$ is 
obtained as $\mu \sim m_{3/2}$.  
But, a value $\mu\sim m_{weak}\ll m_{soft}\sim m_{3/2}$ can be accomodated
for $v_{PQ}< m_{hidden}$, {\it i.e.} if the scale of PQ breaking lies somewhat 
below the mass scale associated with hidden sector SUSY breaking.\footnote{
In models with SUSY breaking arising from {\it e.g.} gaugino condensation
at an intermediate scale $\Lambda_h$, 
then $m_{3/2}\sim \Lambda_h^3/m_P^2$
in which case we would define $m_{hidden}^2 \sim \Lambda_h^3/m_p$.} \footnote{
The model \cite{mafi} shows a more complete ultraviolet theory which includes 
a mechanism to get $v_{PQ}$ in the intermediate scale through the introduction
of a chiral superfield in the hidden brane, yielding an 
ultraviolet suppressed term
in the hidden brane which gives rise to $\mu\sim m_{weak}$ when SUSY is broken 
in the hidden brane through the shining mechanism \cite{nima}.}
A virtue of the KN solution is that it combines a solution to the strong CP
problem with a solution to the SUSY $\mu$ problem which also allows for a 
Little Hierarchy. A further benefit is that it provides an additional
dark matter particle-- namely the DFSZ~\cite{dfsz} axion-- to co-exist with the
(thermally under-produced) higgsino-like WIMP from natural SUSY.
Thus, dark matter is then expected to be comprised of a 
WIMP/axion admixture~\cite{az1,bbc}.
For the lower range of PQ scale $v_{PQ}$, then
the dark matter tends to be axion dominated with typically 10-20\% WIMPs
by mass density~\cite{dfsz2}. For larger $v_{PQ}$ values, then non-thermal processes 
such as saxion and axino~\cite{bci} decay augment the WIMP abundance while for even larger
values of $v_{PQ}$ then the higgsino-like WIMPs are overproduced and one typically
runs into BBN constraints from late-decaying neutral particles 
(saxions and axinos) or overproduction of relativistic axions from saxion decay which contribute to the effective number of neutrino species
$N_{eff}$ 
(which is found to be $N_{eff}=3.13\pm 0.32$ from the recent Particle Data Group tabulation~\cite{pdb}).
In the context of the DFSZ model embedded within the MSSM, 
then the presence of higgsinos in the $a\gamma\gamma$ triangle diagram 
is expected to reduce the axion-photon-photon coupling to levels
below present sensitivity making the SUSY DFSZ axion very challenging to 
detect~\cite{axpaper}.

\subsubsection{Chun-Kim-Nilles}
\label{ssec:ckn}

In the CKN model~\cite{ckn}, it is assumed that SUSY is broken in the hidden sector
due to gaugino condensation 
$\langle \lambda\lambda\rangle\sim \Lambda_h^3\sim (10^{13}$ GeV$)^3$ 
in the presence of a hidden $SU(N)_h$ gauge group. Furthermore, 
there may be vector-like hidden sector quark chiral superfields present
$Q$ and $Q^c$ which transform as $N$ and $N^*$ under $SU(N)_h$. 
The Higgs and hidden quark superfields carry PQ charges as in Table~\ref{tab:ckn}:
\begin{table}[!htb]
\renewcommand{\arraystretch}{1.2}
\begin{center}
\begin{tabular}{c|ccccccc}
multiplet & $H_u$ & $H_d$ & $Q$ & $Q^c$ & $Q_i$ & $U_i^c$ & $D_i^c$ \\
\hline
PQ charge & $-1$ & $-1$  & $1$ & $1$ & 0 & 1 & 1 \\
\hline
\end{tabular}
\caption{PQ charge assignments for various superfields of the CKN model.
}
\label{tab:ckn}
\end{center}
\end{table}
This allows for the presence of a superpotential term
\be
W_{CKN}\ni \frac{\lambda_{\mu}}{m_P}QQ^cH_uH_d .
\ee
Along with gauginos condensing at a scale $\Lambda_h$ to break SUGRA with
$m_{3/2}\sim \Lambda_h^3/m_P^2$, 
the hidden sector scalar squarks condense at a scale $\Lambda <\Lambda_h$
to break the PQ symmetry and to generate a $\mu$ term
\be
\mu_{\rm eff}\sim \lambda_{\mu} \Lambda^2/m_P .
\ee
Thus, this model provides a framework for $\mu <m_{soft}$. 
It also generates a DFSZ axion to solve the strong CP problem
along with a string model-independent (MI) axion which could provide 
a quintessence solution for the cosmological constant (CC)~\cite{Kim:2009cp}.
The CC arises from the very low mass MI axion field slowly settling to the 
minimum of its potential.

\subsubsection{BK/EWK solution linked to inflation and strong CP}
\label{ssec:king}

In Ref's~\cite{BasteroGil:1997vn,EytonWilliams:2004bm}, a model is proposed with superpotential
\be
W_{EWK} \ni \lambda_\mu \phi H_u H_d+\kappa \phi N^2
\ee
where the $\phi$ field plays the role of inflaton and the $N$ field is a
waterfall field leading to hybrid inflation in the early universe~\cite{hybridI}.
Although the model appears similar to the NMSSM, it is based on a 
PQ rather than $\mathbb{Z}_3$ symmetry with charges as in Table \ref{tab:king}. 
Thus, it avoids the NMSSM domain wall problems which arise
from a postulated global $\mathbb{Z}_3$ symmetry. 
Augmenting the scalar potential with soft breaking terms, then the $\phi$ and
$N$ fields gain vevs of order some  intermediate scale $Q\sim 10^{12}$ GeV
so that Yukawa couplings $\lambda_\mu$ and $\kappa$ are of order 
$10^{-10}$. Such tiny Yukawa couplings might arise from type-I 
string theory constructs~\cite{EytonWilliams:2005bg}.
To fulfill the inflationary slow-roll conditions, then the field
$\phi$ must gain a mass of less than $5-10$ MeV and a reheat temperature
of $1-10$ GeV. Domain walls from breaking of the PQ symmetry are inflated away.

\begin{table}[!htb]
\renewcommand{\arraystretch}{1.2}
\begin{center}
\begin{tabular}{c|cccc}
multiplet & $H_u$ & $H_d$ & $\phi$ & $N$ \\
\hline
PQ charge & $-1$ & $-1$  & $+2$ & $-1$  \\
\hline
\end{tabular}
\caption{PQ charge assignments for various superfields 
of the EWK model.
}
\label{tab:king}
\end{center}
\end{table}

\subsubsection{Global symmetries and gravity}
\label{sssec:grav}

It is well known that gravitational effects violate global symmetries, as has been
considered via black hole ``no hair'' theorems~\cite{nohair} and 
wormhole effects~\cite{wormhole}.
In such cases, it has been questioned whether the PQ mechanism can be realistic 
once one includes gravity or embeds the SUSY PQ theory into a 
UV complete string framework~\cite{suss,km_r,dob}.
Indeed, Kamionkowski and March-Russell~\cite{km_r} (KMR) considered the effect of 
gravitational operators such as 
\be
V(\phi )\ni \frac{g}{m_P^{2m+n-4}}|\phi|^{2m}\phi^n+h.c. +c
\ee
involving PQ charged fields $\phi$ in the scalar potential upon the axion potential. In the case of $2m+n=5$, 
{\it i.e.} a term suppressed by a single power of $m_P$, then these gravitational terms
would displace the minimum of the PQ axion potential such that the QCD 
$CP$ violating term $G_{\mu\nu A}\tilde{G}_A^{\mu\nu}$ settles to a non-zero minimum 
thus destroying the PQ solution to the strong CP problem. To maintain 
$\bar{\theta}\alt 10^{-10}$, KMR calculated that all gravitational 
operators contributing to the axion potential should be suppressed by at least
powers of $(1/m_P)^8$. This is indeed a formidable constraint! 

\begin{figure}[tbp]
\centering
\includegraphics[height=0.25\textheight]{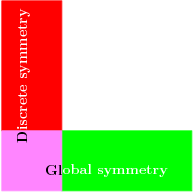}
\caption{Kim diagram~\cite{kn_de,kimrev} where the column represents an 
infinite sequence of Lagrangian terms obeying
gravity-safe discrete symmetry while the row represents an infinite sequence of
terms obeying the global symmetry. The green region terms are gravity-unsafe while
red region violates the global symmetry. The lavender terms are gravity-safe and
obey the global symmetry.
\label{fig:kim}}
\end{figure}

To avoid such terms, additional symmetries are required~\cite{kn3}. 
In string theory, it is known that discrete symmetries arising from gauge symmetries 
are gravity-safe, as are other discrete symmetries or $R$-symmetries arising from
string compactification. 
In Fig. \ref{fig:kim} the Kim diagram is shown~\cite{kn_de,kimrev}. 
The red/lavender column denotes an infinite set of Lagrangian terms in the model 
under consideration which obey some exact, gravity-safe, discrete symmetry. 
Of this set of terms, the few 
lower order terms, denoted by the lavender region, obey an exact global symmetry, 
understood here to be the PQ symmetry whose breaking yields the QCD axion.
The red-shaded terms obey the discrete symmetry but violate any global
symmetry.
The green/lavender row denotes the full, infinite set of global symmetry terms, of which the
green-shaded terms are not gravity-safe. If the discrete symmetry is strong enough, 
then the gravity-unsafe terms will be sufficiently suppressed. 
The global PQ symmetry is expected to be approximate. The question then is: 
is it sufficiently strong so as to be gravity-safe? 
Some additional gravity-safe symmetry is required to ensure the PQ mechanism is robust. 
The lavender region represents gravity-safe terms which obey the global symmetry.

\subsubsection{Gravity-safe symmetries : gauge symmetries or $R$-symmetries: 
continuous or discrete }
\label{sssec:gra_rsym}

Given that global symmetries are not gravity-safe (and hence not fundamental),
it is common to turn to gauge symmetries as a means to forbid the $\mu$ term.
Some models based on an extra local $U(1)^\prime$ were examined in Subsec. \ref{ssec:u1}.
Some problems with this approach emerge in that one has to suitably hide any new
gauge bosons associated with the extra gauge symmetry and one must also typically 
introduce (and hide) extra exotic matter which may be needed to ensure 
anomaly cancellation. In addition, such exotic matter may destroy the desireable feature
of gauge coupling unification should the new exotica not appear in complete GUT multiplets.

An alternative approach is to introduce {\it discrete} gauge 
symmetries~\cite{kn3,CL}. 
Such $\mathbb{Z}_M$ symmetries may emerge from a local $U(1)^\prime$ when a charge 
$M$ object (charged under the new $U(1)^\prime$) condenses at very high
energy leaving a discrete $\mathbb{Z}_M$ gauge symmetry in the low energy effective theory.
Since the $\mathbb{Z}_M$ emerges from a local gauge theory, it remains gravity-safe.
In Subsec. \ref{ssec:mbgw}, the MBGW model~\cite{bgw2} which is based on a 
$\mathbb{Z}_{22}$ discrete gauge symmetry is examined. 
The model under $\mathbb{Z}_{22}$ is found to be anomaly-free and is used to not only
forbid the $\mu$ term but to generate a PQ symmetry needed to solve the strong CP problem.
The lowest order PQ violating term allowed by the $\mathbb{Z}_{22}$ 
is sufficiently suppressed so that PQ arises as an  
accidental approximate global symmetry thereby rendering the model to be gravity-safe. 
The $\mathbb{Z}_{22}$ discrete gauge charges of the multiplets turn out to be 
not consistent with GUTs which should be manifested at some level in the high energy theory. 
Also, the presence of a charge 22 object which  condenses
at some high energy scale may not be very plausible and might be inconsistent 
with the UV completion of the theory ({\it i.e.} lie in the swampland). 

Continuous or discrete $R$-symmetries offer a further choice for gravity-safe symmetries.
A solution using a continuous $U(1)_R$ symmetry was examined in Subsec. \ref{ssec:Rsym}.\footnote{See also Ref. \cite{Choi:2010xf}.}
In the interest of minimality, it is noted that continuous $R$ symmetries
are not consistent with the particle content of just the MSSM~\cite{dreiner}.
Then it is also of interest to examine the possibility of discrete 
remnant $R$-symmetries $\mathbb{Z}_N^R$ which arise upon compactification
of the full Lorentz symmetry of 10-$d$ string theories.
$R$-symmetries are characterized by the fact that superspace co-ordinates 
$\theta$ carry non-trivial $R$-charge: 
in the simplest case, $Q_R(\theta )=+1$ so that $Q_R( d^2\theta ) =-2$. 
For the Lagrangian ${\cal L}\ni \int d^2\theta W$ to be invariant under 
$\mathbb{Z}_N^R$-symmetry, the superpotential $W$ must carry $Q_R(W)= 2 + $integer 
multiples of $N$. 

\begin{table}[!htb]
\renewcommand{\arraystretch}{1.2}
\begin{center}
\begin{tabular}{c|ccccc}
multiplet & $\mathbb{Z}_{4}^R$ & $\mathbb{Z}_{6}^R$ & $\mathbb{Z}_{8}^R$ & $\mathbb{Z}_{12}^R$ & $\mathbb{Z}_{24}^R$ \\
\hline
$H_u$ & 0  & 4  & 0 & 4 & 16 \\
$H_d$ & 0  & 0  & 4 & 0 & 12 \\
$Q$   & 1  & 5 & 1 & 5  & 5 \\
$U^c$ & 1  & 5 & 1 & 5  & 5 \\
$E^c$ & 1  & 5 & 1 & 5  & 5 \\
$L$   & 1  & 3 & 5 & 9  & 9 \\
$D^c$ & 1  & 3 & 5 & 9  & 9 \\
$N^c$ & 1  & 1 & 5 & 1  & 1 \\
\hline
\end{tabular}
\caption{Derived MSSM field $R$ charge assignments for various anomaly-free 
discrete $\mathbb{Z}_{N}^R$ symmetries which are consistent with $SU(5)$ or 
$SO(10)$ unification (from Lee {\it et al.} Ref.~\cite{lrrrssv2}).
}
\label{tab:R}
\end{center}
\end{table}

These remnant discrete $R$-symmetries $\mathbb{Z}_N^R$-- if sufficiently strong--
can forbid lower order operators in powers of $1/m_P$ which would violate
putative global symmetries such as PQ.
Such a built-in mechanism from string theory may enable the PQ symmetry 
to be strong enough to support the axion solution
to the strong CP problem. Since the $R$-symmetry is necessarily supersymmetric 
(it acts on superspace co-ordinates), this is another instance in how the implementation of 
the axion solution to the strong CP problem is enhanced and made more plausible by the 
presence of supersymmetry. 
However, not all possible $R$-symmetries are a suitable candidate for a fundamental 
symmetry. 
Table \ref{tab:R} (as derived in Ref's~\cite{lrrrssv1,lrrrssv2}) 
shows the $R$-symmetries along with the $R$-charges of the multiplets 
which are consistent with either $SU(5)$ or 
$SO(10)$ unification, anomaly-free (allowing for a Green-Schwarz term),
forbid the $\mu$ term and also forbid the 
R-parity violating and dimension-five proton decay operators and 
hence can serve the purpose of being a fundamental symmetry.
In fact, the  $\mathbb{Z}_N^R$ symmetries of Table \ref{tab:R} have been shown to be the
{\it only} anomaly-free symmetries which allow for fermion masses and suppress the
$\mu$ term while maintaining consistency with GUTs. As a bonus, they allow for
neutrino masses while forbidding $R$-parity and dangerous proton decay operators. 
Implementation of the discrete $R$-symmetries is only possible in extra-dimensional
GUTs, making their implementation in string compactifications very natural~\cite{cfr}.

\subsubsection{Natural Higgs-Flavor-Democracy (HFD) solution to $\mu$ problem}
\label{sssec:hfd}

%\KJB{I am not sure that this is a good example. 
%It is quite unclear how this model protects light QCD axion:
%we may need more explanation for derivation of Eq.~\eqref{eq:HFD4} 
%and need to show that other potential PQ violating terms are suppressed or absent.}

In Ref.~\cite{jkim}, the $\mu$ problem is solved by introducing additional identical 
Higgs doublet superfields to those of the MSSM. 
The theory then contains a direct product of discrete interchange symmetries 
$S_{2}(H_u)\times S_{2}(H_d)$. 
This is {\it Higgs flavor democracy} (HFD).
Besides solving the $\mu$ problem, this mechanism also gives rise to an approximate PQ symmetry and 
hence a light QCD axion, thereby solving the strong CP problem whilst avoiding the 
gravity spoliation problem. 
The HFD discrete symmetry can be found in several string theory models.

\textit{HFD:} One starts by introducing two pairs of Higgs doublets 
at the GUT scale $m_G$ namely : \{$H_u^{(1)}$, $H_d^{(1)}$\} and \{$H_u^{(2)}$, $H_d^{(2)}$\}. 
However, the weak scale MSSM requires only one pair of Higgs doublets: \{$H_u$, $H_d$\}. 
If, at the GUT scale, the two pairs of Higgs doublets : 
$H_u$ = \{$H_u^{(1)}$, $H_u^{(2)}$\} and $H_d$ = \{$H_d^{(1)}$, $H_d^{(2)}$\} 
are indistinguishable then there must exist the permutation
symmetries $S_2(H_u)\times S_2(H_d)$ . 
Then the Higgsino mass matrix has a democratic form given by: 
\[ 
\left \{
  \begin{tabular}{cc}
  $m_G$/2 & $m_G$/2  \\
  $m_G$/2 & $m_G$/2 \\
  \end{tabular}
\right \} .
\] 
The Higgs mass eigenvalues are $m_G$ and 0. 
Hence, the Higgs pair in the weak scale MSSM is obtained to be massless. 
Still, the model construction of the MSSM requires a massive Higgs pair at the weak scale
with mass value $\mu$. 
In order to fulfill this criteria, the HFD must be broken and this mechanism 
results in $\mu$ $\approx$ $\mathcal{O}$(TeV).

\textit{Generation of $\mu$:} The minimal Kahler potential is considered as $K = \Phi_i \Phi_i^{\dagger}$ 
where $\Phi_i$ ( i =1, 2) is a doublet under the gauge group such as the Higgs superfield and 
$X_i$ and $\bar{X_i}$ (i=1,2) are singlets under the gauge group. 
Both $\Phi_i$ and $X_i$ and the corresponding barred fields obey the $S_2\times S_2$ symmetry. 
$X^{(0)}$ and $\bar{X}^{(0)}$ are SM singlet fields containing a very light QCD axion for 
$10^9$ GeV $\leq$ $v_{PQ}$ $\leq$ $10^{12}$ GeV. 
With this construct, the $S_2(L) \times  S_2(R)$ symmetric nonrenormalizable term is:
%\KJB{Notations are not consistent. What is $M_p'$?}
\begin{equation}
    W^{(nonrenormalizable)} = \sum_{i,j=1,2} \Bigg( \frac{X^{(i)}\bar{X}^{(j)}}{m_P} \Bigg) H_u^{(i)} H_d^{(j)}
    + \sum_{ij} \sum_{kl} \Bigg( \frac{X^{(i)}\bar{X}^{(j)}}{m_P} \Bigg) H_u^{(k)} H_d^{(l)}
\label{eq:HFD2}
\end{equation}
With the HFD breaking minimum at $\langle X_1 \rangle$ = $\langle \bar{X_1} \rangle$ = $v_{PQ}$ 
and $\langle X_2 \rangle$ = $\langle \bar{X_2} \rangle$ = 0, Eq. \eqref{eq:HFD2} becomes 
\begin{equation}
    W^{(nonrenormalizable)} = \frac{\lambda_\mu  v_{PQ}^2}{2m_P}(H_u^{(0)}+H_u^{(M_G)})(H_d^{(0)}+H_d^{(M_G)})
\label{eq:HFD3}
\end{equation}
This choice of HFD breaking minimum is spontaneous.
Thus we obtain $\mu$ = $\frac{\lambda_\mu  v_{PQ}^2}{2m_P}$. 
With $10^{10}$ GeV $\leq$ $v_{PQ}$ $\leq$ $10^{12}$ GeV and $\lambda_\mu$ $\approx$ $\mathcal{O}$(1), 
we obtain $\mu$ $\approx$ $\mathcal{O}(0.1-10^3 \text{ TeV})$. 
The LH can be accomodated for the lower range of $v_{PQ}$ or if $\lambda_\mu <1$.

\textit{Light QCD Axion} - Integrating out the heavy fields in Eq. \eqref{eq:HFD3}, one obtains
\begin{equation}
    W = \frac{\lambda_\mu  X^{(0)} \bar{X}^{(0)}}{2m_P}H_u^{(0)}H_d^{(0)} .
\label{eq:HFD4}
\end{equation}
The PQ charges of Higgs multiplets are obtained from their interaction with the 
quarks and PQ charges of $X^{(0)}$ and $\bar{X}^{(0)}$ are defined by Eq. \eqref{eq:HFD4}. 
Thus, a term $m_{3/2}\frac{\lambda^2}{4m_P^2}\frac{1}{M_G}H_uH_d(XX^c)^2$ is 
obtained which violates PQ and hence adds a tiny correction to $\mu$. 
Here, $M_G$ is the GUT scale higgsino mass. 
Hence, PQ symmetry emerges as an approximate symmetry, thereby giving rise to a light QCD axion 
which does not suffer from the gravity-spoliation problem.

\subsubsection{Radiative PQ breaking from SUSY breaking}
\label{ssec:radpq}

The above models are particularly compelling in that
they include supersymmetry which solves the gauge hierarchy problem, 
but also include the axion solution to the strong CP problem of QCD. 
In addition, they allow for the required Little Hierarchy of $\mu\ll m_{soft}$.
A drawback to the KN model is that it inputs the PQ scale ``by hand'' 
via the superpotential Eq.~\eqref{eq:knW}. It is desireable if the PQ scale can 
be generated via some mechanism and furthermore, the emergence of
three {\it intermediate mass scales} in nature-- the hidden sector
SUSY breaking scale, the PQ scale and the Majorana neutrino scale-- 
begs for some common origin. 
A model which accomplishes this was first proposed
by Murayama, Suzuki and Yanagida (MSY)~\cite{msy}.

In radiative PQ breaking models, the MSSM superpotential is
\be
W_{MSSM}=\sum_{i,j=1}^{3}\left[ ({\bf f}_u )_{ij}Q_iH_u U_j^c+ 
({\bf f}_d )_{ij}Q_iH_d D_j^c+({\bf f}_e )_{ij}L_iH_d E_j^c+({\bf f}_{\nu} )_{ij}
L_iH_u N_j^c \right]
\ee
where we explicitly include the right hand neutrino superfields $N_i$ and the
generation indices $i,j$ run from $1-3$. 
To this, we add a PQ superpotential containing new PQ-charged fields $X$ and $Y$ of
the form
\be
W_{PQ}\ni \frac{1}{2}h_{ij}XN_i^cN_j^c+\frac{f}{m_P}X^3Y+W_{\mu}
\label{eq:Wmsy}
\ee
and where 
\be
W_{\mu}^{MSY} =\frac{g_{MSY}}{m_P}XYH_uH_d ,
\label{eq:msy}
\ee

where the PQ charges $Q_{PQ}(matter)=1/2$, $Q_{PQ}(Higgs)=-1$, $Q_{PQ}(X)=-1$
and $Q_{PQ}(Y)=3$. 
Along with the MSY superpotential terms, we include the corresponding 
soft SUSY breaking terms
\bea
V_{MSY}& \ni & m_X^2|\phi_X|^2+m_Y^2|\phi_Y|^2+m_{N_i}^2|\phi_{N_i}|^2\nonumber \\
&+& \left( \frac{1}{2}h_iA_i\phi_{N_i}^2\phi_X+\frac{f}{m_P}A_f\phi_X^3\phi_Y+
\frac{g_{MSY}}{m_P}A_gH_uH_d\phi_X\phi_Y +h.c.\right).
\eea
For simplicity, we assume a diagonal coupling $h_{ij}= h_i\delta_{ij}$.
The model may be defined as applicable at the reduced 
Planck scale $m_P\simeq 2.4\times 10^{18}$ GeV and the corresponding 
Renormalization Group Equations (RGEs) can be found in Ref.~\cite{msy}
at 1-loop and Ref.~\cite{radpq} at 2-loop order.
Under RG evolution, the large Yukawa coupling(s) $h_i$ push the 
soft mass $m_X^2$ to negative values at some intermediate mass scale resulting
in the radiatively-induced breakdown of PQ symmetry as a consequence of 
SUSY breaking. 
The scalar potential consists of the terms $V=V_F+V_D+V_{\rm soft}$.
The Higgs field directions can be ignored since these develop vevs at much lower
energy scales. Then the relevant part of the scalar potential is just
\be
V_F\ni \frac{|f|^2}{m_P^2}|\phi_X^3|^2+\frac{9|f|^2}{m_P^2}|\phi_X^2\phi_Y|^2 .
\ee
Augmenting this with $V_{\rm soft}$, we minimize $V$ at a scale $Q=v_{PQ}$ 
to find the vevs of $\phi_X$ and $\phi_Y$ ($v_X$ and $v_Y$):
\bea
0&=& \frac{9|f|^2}{m_P^2}|v_X^2|^2v_Y +f^*\frac{A_f^*}{m_P}v_X^{*3}+m_Y^2v_Y 
\label{eq:minQ}\\
0&=& \frac{3|f|^2}{m_P^2}|v_X^2|^2v_X+\frac{18|f|^2}{m_P^2}|v_X|^2|v_Y|^2v_X
+3f^*\frac{A_f^*}{m_P}v_X^{*2}v_Y^*+m_X^2v_X .
\label{eq:minP}
\eea
The first of these may be solved for $v_Y$. Substituting into the second, we find a 
polynomial for $v_X$ which may be solved for numerically. 
The potential has two minima in the $v_X$ and $v_Y$ plane 
symmetrically located with respect to the origin. 
For practical purposes, we use the notation $v_X$=$|v_X|$ and $v_Y$=$|v_Y|$.

\begin{figure}[tbp]
\centering
\includegraphics[height=0.4\textheight]{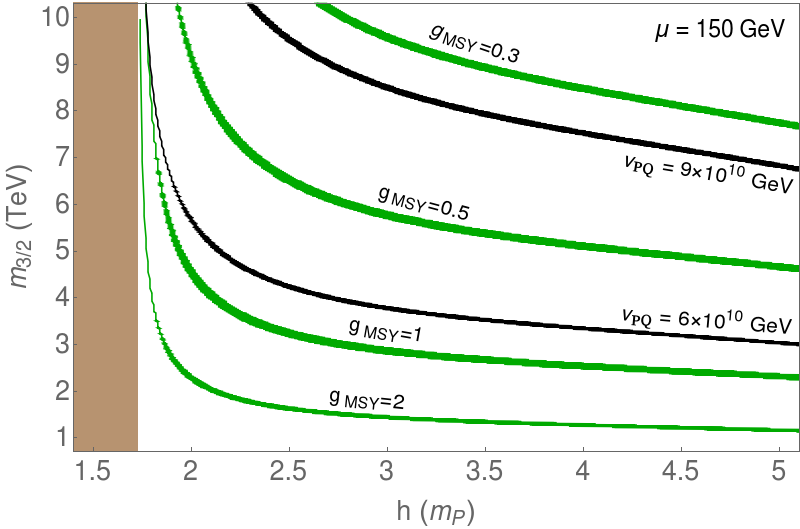}
\caption{Value of $g$ which is needed in the MSY to generate $\mu =150$ GeV 
from a gravitino mass $m_{3/2}$ and a GUT coupling $h$.
We also show some contours of $v_{PQ}$.
\label{fig:gMSY}}
\end{figure}

The fields $\phi_X$ and $\phi_Y$ obtains vevs $v_X$ and $v_Y$
at the intermediate mass scale, taken here to be $v_{PQ}=\sqrt{v_X^2+9v_Y^2}$.
The corresponding axion decay constant is given by $f_a = \sqrt{2}v_{PQ}$.\footnote{For axion interactions,
the actual decay constant is $f_A\equiv f_a/N_{DW}$ where $N_{DW}$ is the domain wall number.} 
%\KJB{I redefined $v_{PQ}$ and $f_a$ for consistency. 
%I don't think we have to include $f_a$ lines in Fig.~\ref{fig:gMSY}, 
%\ref{fig:gCCK} and \ref{fig:gSPM}. 
%Please change contours and numbers in those figures with new $v_{PQ}$ definition.}
A DFSZ-like axion $a$ arises as the pseudo-Goldstone boson of spontaneous
PQ breaking, thus solving the strong CP problem. A $\mu$ parameter, which is
originally forbidden by PQ symmetry, is generated with a value
\be
\mu_{\rm eff}=g_{MSY}\frac{v_Xv_Y}{m_P}
\ee
and a Majorana neutrino mass, also initially forbidden by PQ symmetry, 
is generated at 
\be
M_{N_i}=h_i|_{Q=v_x} v_X .
\ee
Since the $\mu$ term depends on an arbitrary coupling $g_{MSY}$, one may obtain any desired
value of $\mu$ for particular $v_X$ and $v_Y$ vevs by suitably adjusting $g_{MSY}$. 
However, if the required values of $g_{MSY}$ are very different from unity, 
{\it i.e.} $g_{MSY}\gg 1$ or $g_{MSY}\ll 1$, 
we might need to introduce an additional physical scale to explain the $\mu$ term.
To generate a value of $\mu =150$ GeV, then values of $g_{MSY}$ as shown in Fig. \ref{fig:gMSY} 
are required depending on the values of $m_{3/2}$ and $h(M_P)$ which are assumed.

The virtues of this model then include:
\begin{itemize}
\item it is supersymmetric, thus stabilizing the Higgs sector and allowing for
a gauge hierarchy,
\item it solves the strong CP problem via a DFSZ-like axion $a$,
\item it presents a unified treatment of the three intermediate mass scale
where the PQ and Majorana neutrino scales arise as a consequence of 
SUSY breaking and
\item it allows for a Little Hierarchy $\mu\ll m_{soft}$ for the case where
$v_{PQ}< m_{hidden}$.
\end{itemize}

Detailed numerical calculations in the MSY model have been carried out in Ref.~\cite{radpq}. 
There, it is found that for generic $W_{\mu}^{MSY}$ couplings
$g_{MSY}\sim 0.1-1$, then a $\mu$ parameter $\mu\sim 100-200$ GeV can easily be 
generated from TeV-scale soft breaking terms.
Furthermore, since the $\mu$ term sets the mass scale for the $W,Z,h$ 
boson masses and is determined itself by the PQ vevs $v_X$ and $v_Y$, then
the axion mass 
$m_a\simeq 0.48 f_{\pi}m_{\pi}/f_a=6.25\times 10^{-3}\ {\rm GeV}/f_a$ 
is related to the Higgs mass $m_h$ and the higgsino masses 
$m_{\tw_1,\tz_{1,2}}\sim \mu$.
%Due to the several beneficial properties of the MSY model detailed above, 
%we investigate further numerics of this and related models in 
%Sec. \ref{sec:radpq}.
%{\bf Discuss numerics of $f_a, m_{N_i}$.}
The required PQ charges for the MSY model are listed in Table \ref{tab:radpq}.
\begin{table}[!htb]
\renewcommand{\arraystretch}{1.2}
\begin{center}
\begin{tabular}{c|ccc}
multiplet & MSY & CCK & SPM  \\
\hline
$H_u$ & $-1$   & $-1$  & $-1$ \\
$H_d$ & $-1$   & $-1$  & $-1$ \\
$Q$   & $+1/2$ & $3/2$  & $+1/2$ \\
$L$   & $+1/2$ & $3/2$  & $+5/6$ \\
$U^c$ & $+1/2$ & $-1/2$ & $+1/2$ \\
$D^c$ & $+1/2$ & $-1/2$ & $+1/2$ \\
$E^c$ & $+1/2$ & $-1/2$ & $+1/6$  \\
$N^c$ & $+1/2$ & $-1/2$ & $+1/6$ \\
$X$   &  $-1$  & $+1$   & $-1/3$  \\
$Y$   & $+3$   & $-3$   & $+1$ \\
\hline
\end{tabular}
\caption{PQ charge assignments for various superfields of the CCK, MSY and SPM
models of radiative PQ breaking. 
}
\label{tab:radpq}
\end{center}
\end{table}

\begin{figure}[tbp]
\centering
\includegraphics[height=0.4\textheight]{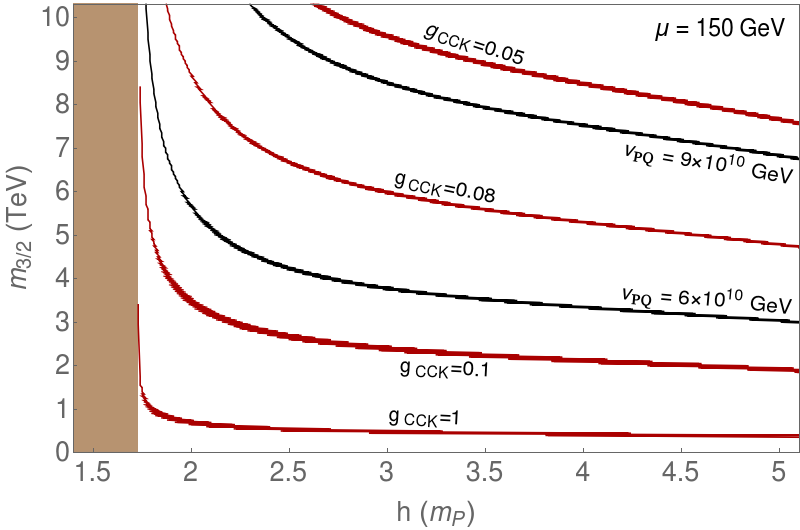}
\caption{Value of $g$ which is needed in the CCK to generate $\mu =150$ GeV 
from a gravitino mass $m_{3/2}$ and a GUT coupling $h$.
We also show some contours of $v_{PQ}$.
\label{fig:gCCK}}
\end{figure}
%
%
\begin{figure}[tbp]
\centering
\includegraphics[height=0.4\textheight]{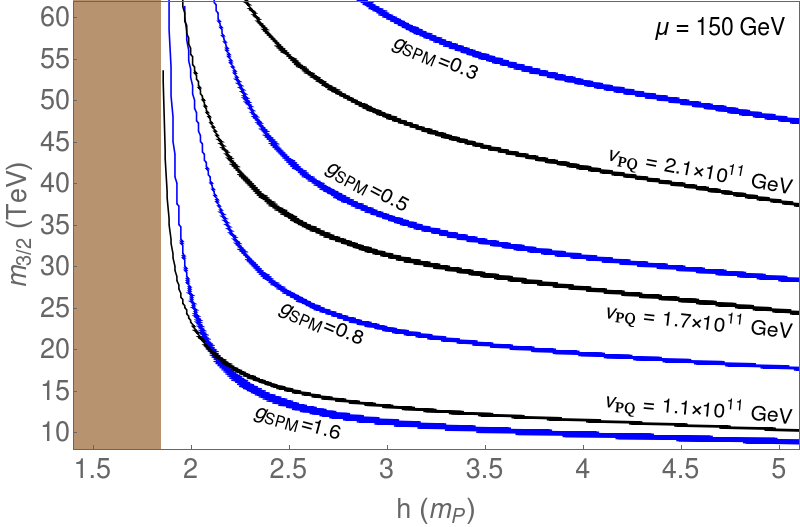}
\caption{Value of $g$ which is needed in the SPM model to generate $\mu =150$ GeV 
from a gravitino mass $m_{3/2}$ and a GUT coupling $h$.
We also show some contours of $v_{PQ}$.
\label{fig:gSPM}}
\end{figure}
%

%
%\begin{figure}[tbp]
%\includegraphics[height=0.4\textheight]{g_un}
%\caption{Value of $g$ which is needed in the CCK, MSY and SPM models 
%to generate $\mu =150$ GeV from a gravitino mass $m_{3/2}$ and a GUT coupling $h$.
%\label{fig:g_mu150}}
%\end{figure}
%

Other closely related models make different choices for which fields enter into $W_{\mu}$.
We can also have:
\bea
W_{\mu}^{CCK} &=&\frac{g_{CCK}}{m_P}X^2 H_uH_d\ \ \ \ or\ \\
W_{\mu}^{SPM} &=&\frac{g_{SPM}}{m_P}Y^2 H_uH_d .
\eea
The above three possibilities for $W_{\mu}$ correspond to Ref's~\cite{msy} (MSY), 
\cite{cck} (CCK) and \cite{spm} (SPM).
The corresponding PQ charges for the three radiative PQ breaking models are listed in 
Table \ref{tab:radpq}.

We list in Fig's \ref{fig:gCCK} and \ref{fig:gSPM} also the values of $g_{CCK}$ and $g_{SPM}$ which are needed
to generate a value of $\mu\simeq 150$ GeV. For a given value of $h(m_P)$ and $m_{3/2}$, 
then typically $g_{CCK}< g_{MSY}<g_{SPM}$. 
The MSY model has the interesting
feature that the PQ charge assignments are consistent with $SO(10)$ unification. 
We also remark that all three models can easily generate
weak scale values of $\mu$ from multi-TeV values of $m_{3/2}$: 
{\it i.e.} $\mu\ll m_{3/2}$ so that a Little Hierarchy is naturally generated.

{\it Gravity safety of radiative PQ breaking models:} 
An important issue for the radiative PQ breaking models is whether the 
required PQ symmetry is actually gravity-safe and whether it may emerge from
any of the aforementioned $\mathbb{Z}_N^R$ symmetries.
We have examined whether or not the three radiative PQ breaking models of 
Table \ref{tab:radpq} (CCK, MSY and SPM) can be derived from any of the more
fundamental $\mathbb{Z}_N^R$ symmetries in Table \ref{tab:R}~\cite{grav}.
In almost all cases, the $hXN^cN^c$ operator is disallowed: then there is no
large Yukawa coupling present to drive the PQ soft term $m_X^2$ negative 
so that PQ symmetry is broken. And since the PQ symmetry does not allow for
a Majorana mass term $\frac{1}{2}M_NN^cN^c$, then no see-saw scale can be developed.
One exception is the MSY model under $\mathbb{Z}_4^R$ symmetry with charge
assignments $Q_R(X)=0$ and $Q_R(Y)=2$: then a $YH_uH_d$ term is allowed
which would generate a $\mu$ term of order the intermediate scale.
Also, without considering any specific R-charges for the fields $X$ and $Y$, 
we can see that the R-charges for $X$ and $Y$ should be such that the term 
$XYH_uH_d$ is allowed and since the R-charges of $H_u$ and $H_d$ are 0, 
then a term  $MXY$ would always be allowed: this term breaks PQ 
at high order and is not gravity safe.
A second exception is SPM under the $\mathbb{Z}_6^R$ symmetry with
charges $Q_R(X)=0$ and $Q_R(Y)=2$: then operators like $Y^4/m_p$ are allowed
which break PQ but are not sufficiently suppressed so as to be gravity-safe. 
Furthermore, we can see that in this model that the R-charge of $Y$ 
is such that terms like  $M^2Y$ which break PQ are
always allowed but are not gravity safe.
Thus, we conclude that while the radiative PQ breaking models are indeed compelling and
can address all three intermediate scales in a unified framework, the 
required PQ symmetry does not appear gravity-safe.

\subsubsection{CCL model from gauged $U(1)_R$ symmetry}
\label{ssec:CCL}

In the model of Choi, Chun and Lee~\cite{Choi:2010xf} (CCL), 
the $\mu$ term is generated in a manner similar to the SPM model~\cite{spm}, 
but with the difference that the fundamental symmetry is a 
gauged $U(1)_R$ symmetry out of which the PQ symmetry arises to be an 
accidental approximate symmetry. 
The superpotential for CCL is
\bea
W_{CCL}&=&f_uQH_uU^c+f_dQH_dD^c+f_eLH_dE^c+f_{\nu}LH_uN^c+\\
&+&\lambda_{\mu}\frac{Y^2H_uH_d}{m_p}+\kappa X^3Y/m_P+\lambda_NX^nN^cN^c/2m_P^{n-1} ,
\label{eq:WCCL}
\eea
with $U(1)_R$ and $PQ$ charges for the $n=2$ case given in Table~\ref{tab:CCL}.
\begin{table}[!htb]
\renewcommand{\arraystretch}{1.2}
\begin{center}
\begin{tabular}{c|cccccccccc}
multiplet & $H_u$ & $H_d$ & $Q_i$ & $L_i$ & $U_i^c$ & $D_i^c$ & $E_i^c$ & $N_i^c$ & X & Y \\
\hline
$U(1)_R$ charge & $4$ & $4$  & $-\frac{4}{3}$ & $-\frac{4}{3}$ 
& $-\frac{2}{3}$& $-\frac{2}{3}$ & $-\frac{2}{3}$ & $-\frac{2}{3}$ 
& $\frac{5}{3}$ & $-3$ \\
\hline
PQ charge & $3$ & $3$  & $-3$ & $-2$ & $0$ & $0$ & $-1$ & $-1$ & $1$ & $-3$ \\
\hline
\end{tabular}
\caption{$U(1)^R$ and PQ charge assignments for various superfields of the 
CCL model for $n=2$.
}
\label{tab:CCL}
\end{center}
\end{table}

The singlets $X$ and $Y$ get their VEVs at the intermediate scale 
when the PQ symmetry is broken via a large (relative to $m_{3/2}$) 
negative trilinear soft term contribution to the scalar potential, 
thereby giving rise to $\mu\sim m_{soft}$. 
The $U(1)_R$ gauge boson has mass of order the compactification scale
so the low energy theory is that of the MSSM.
Because the fundamental symmetry of CCL is a gauged $U(1)_R$ symmetry, 
the phenomenology of this model is dictated by a hierarchy of soft terms 
$m_{1/2}\gg m_{scalars}>m_{3/2}$ ($m_{1/2}$: gaugino mass). 
Scalar soft masses are fixed in terms of $U(1)_R$ $D$-terms
and typically lead to large negative $m_{H_u}^2$ at the weak scale
which then requires a large, unnatural $\mu$ term which would violate 
the $\mu\ll m_{soft}$ Little Hierarchy.
The gravitino or the RH sneutrino turns out to be the LSP and hence 
end up as cold dark matter candidates. 
If the neutrino is Majorana type then the gravitino is the LSP and 
if the neutrino is Dirac type then the RH sneutrino is the LSP.

\subsubsection{MBGW model of PQ breaking from SUSY breaking}
\label{ssec:mbgw}

\begin{figure}[tbp]
\centering
\includegraphics[height=0.4\textheight]{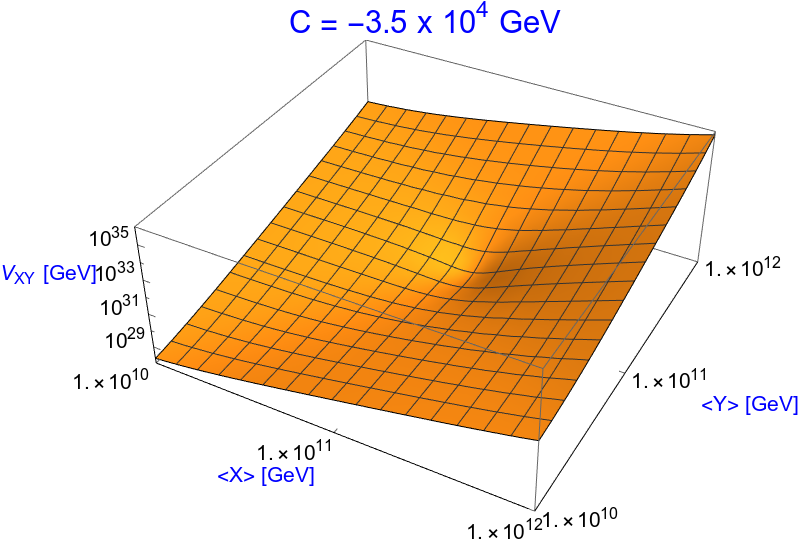}
\caption{Scalar potential $V_{MBGW}$ versus $\phi_X$ and $\phi_Y$ for 
$m_s=10^4$ GeV and $C=-3.5\times 10^4$ GeV.
\label{fig:Vbgw}}
\end{figure}

The Martin-Babu-Gogoladze-Wang (MBGW) model~\cite{spm,bgw2} begins with a 
superpotential
\bea
W&=&f_uQH_uU^c+f_dQH_dD^c+f_eLH_dE^c+f_{\nu}LH_uN^c\\
&&+\frac{1}{2}M_RN^cN^c+\lambda_{\mu}\frac{X^2H_uH_d}{m_p}+\lambda_2\frac{(XY)^2}{m_P}
\label{eq:mbgw}
\eea
which is augmented by soft SUSY breaking terms
\be
V_{soft}\ni m_X^2|\phi_X|^2+m_Y^2|\phi_Y|^2+\left(\lambda_2C\frac{(\phi_X\phi_Y)^2}{m_P}
+h.c.\right)
\ee
so that the scalar potential is
\be
V_{MBGW}=V_F+V_{soft}
\ee
with 
\be
V_F\ni 4\frac{\lambda_2^2}{m_P}|\phi_X\phi_Y|^2\left( |\phi_X|^2+|\phi_Y|^2\right).
\ee
The scalar potential admits non-zero minima in the fields $\phi_X$ and $\phi_Y$ for
$C<0$. The scalar potential for the case of $m_X=m_Y\equiv m_s=10^4$ GeV 
and $C=-3.5\times 10^4$ GeV is shown in Fig. \ref{fig:Vbgw}.

It is found in Ref.~\cite{bgw2} that the model admits a remnant $\mathbb{Z}_{22}$ 
discrete gauge symmetry which is anomaly free up to Green-Schwarz terms and forbids 
lower order operators which would lead to gravitational instability. 
Beside the terms in Eq.~\eqref{eq:mbgw}, the lowest order 
PQ-violating term in the superpotential is $\frac{(Y)^{11}}{m_P^8}$: 
thus this model is gravity safe according to the KMR criterion.
An approximate PQ symmetry emerges as an accidental consequence of the discrete 
$\mathbb{Z}_{22}$ gauge symmetry.
The $\mathbb{Z}_{22}$ and PQ charges are listed in Table \ref{tab:bgw}.
\begin{table}[!htb]
\renewcommand{\arraystretch}{1.2}
\begin{center}
\begin{tabular}{c|cccccccccc}
multiplet & $H_u$ & $H_d$ & $Q_i$ & $L_i$ & $U_i^c$ & $D_i^c$ & $E_i^c$ & $N_i^c$ & $X$ & $Y$ \\
\hline
$\mathbb{Z}_{22}$ charge & $22$ & $18$  & $3$ & $11$ & $19$ & $1$ & $15$ & $11$ & $13$ & $20$ \\
\hline
PQ charge & $-1$ & $-1$  & $+1$ & $+1$ & $0$ & $0$ & $0$ & $0$ & $+1$ & $-1$ \\
\hline
\end{tabular}
\caption{$\mathbb{Z}_{22}$ and PQ charge assignments for various superfields of the 
MBGW model.
}
\label{tab:bgw}
\end{center}
\end{table}

\begin{figure}[tbp]
\centering
\includegraphics[height=0.4\textheight]{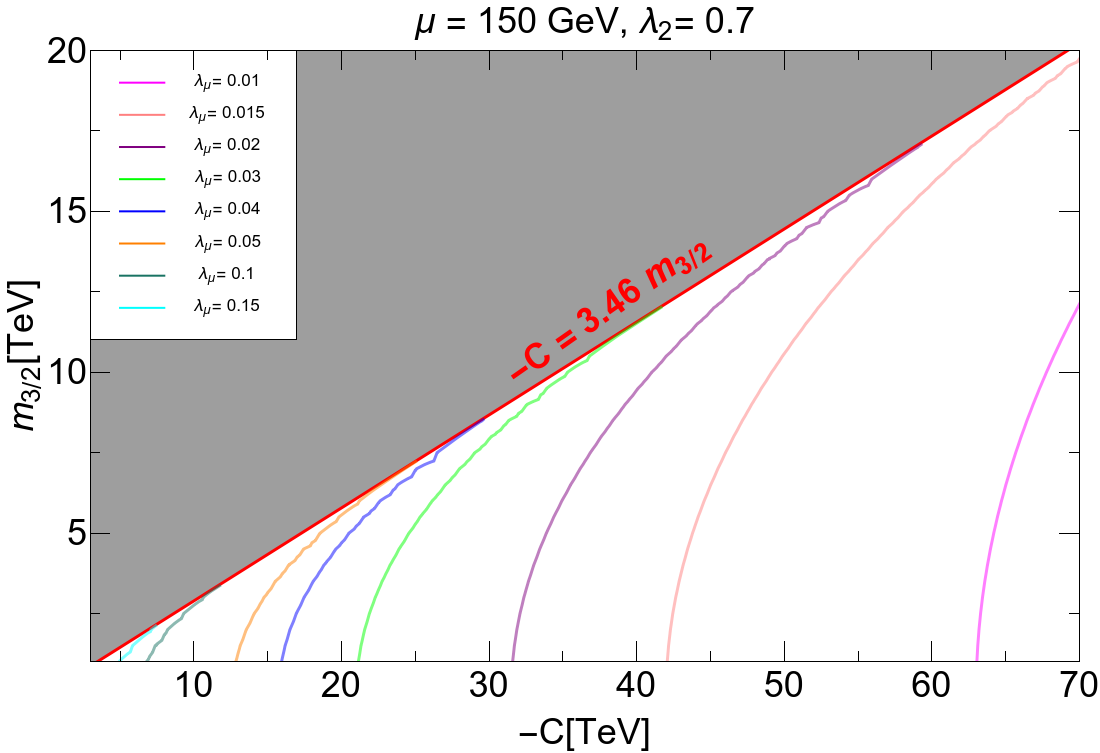}
\caption{Value of $\lambda_{\mu}$ required for $\mu =150$ Gev 
in the $m_{3/2}$ vs. $-C$ plane of the MBGW model.
\label{fig:bgw}}
\end{figure}

By taking $\langle \phi_X\rangle\equiv v_x$ and $\langle\phi_Y\rangle \equiv v_Y$, 
then the scalar potential minimization conditions read
\bea
0 &=& 
2\frac{\lambda_2}{m_P}C^*v_xv_Y^2+m_X^2v_X+4\frac{\lambda_2^2}{m_P^2}\left(
v_Xv_Y^2(v_X^2+v_Y^2)+v_X^3v_Y^2\right)\\
0 &=& 
2\frac{\lambda_2}{m_P}C^*v_x^2v_Y+m_Y^2v_Y+4\frac{\lambda_2^2}{m_P^2}\left(
v_X^2v_Y(v_X^2+v_Y^2)+v_X^2v_Y^3\right).
\eea
A simplifying assumption of $m_X^2=m_Y^2\equiv m_s^2$ and $v_X=v_Y\equiv v_s$ leads to
\be
v_s^2=\frac{-C\pm\sqrt{C^2-12m_s^2}}{12\lambda_2}m_P
\ee
so that the $\mu$ term is
\be
\mu_{MBGW}\simeq\lambda_\mu\frac{v_s^2}{m_P}
\ee
with $v_s^2\simeq \frac{|C|}{12\lambda_2}m_P$.
Taking $m_s\simeq m_{3/2}=10^4$ GeV with $\mu =150$ GeV and 
$C=-3.5\times 10^4$ GeV
leads to $v_s\simeq v_{PQ}\simeq 10^{11}$ GeV for $\lambda_2=0.7$ and 
$\lambda_{\mu}\simeq0.036$. Thus, the MBGW model admits a Little Hierarchy 
$\mu\ll m_{3/2}$ whilst generating the PQ scale $v_{PQ}\sim 10^{11}$ GeV 
(which generates mainly axion dark matter with a smaller portion of higgsino-like WIMPs~\cite{bbc,dfsz2,axpaper}).
The allowed range of MBGW model parameter space is shown in Fig. \ref{fig:bgw}
where we show contours of $\lambda_{\mu}$ values which lead to $\mu =150$ GeV.

As mentioned previously, the MBGW model appears gravity-safe under the $\mathbb{Z}_{22}$ 
discrete gauge symmetry,
The discrete gauge symmetry $\mathbb{Z}_M$ might arise if a charge $Me$ 
field condenses and is integrated out of the low energy theory while
charge $e$ fields survive (see Krauss and Wilczek, Ref.~\cite{grav}). 
While the ensuing low energy theory should be gravity safe, 
for the case at hand one might wonder at the 
plausibility of a condensation of a charge 22 object and whether it might 
occupy the so-called {\it swampland}~\cite{swamp} of theories not 
consistent with a UV completion in string theory. 
In addition, the charge assignments~\cite{bgw2} are not consistent with  
$SU(5)$ or $SO(10)$ grand unification which may be expected at some
level in a more ultimate theory.

Alternatively, it is worth checking whether MBGW is gravity-safe under any of the 
discrete $R$-symmetries listed in Table \ref{tab:R}.
To check gravity safety, we note that additional superpotential terms of the
form $\lambda_3X^pY^q$ may be allowed for given $\mathbb{Z}_N^R$ charge 
assignments and powers $p$ and $q$. Such terms will typically break the PQ
symmetry and render the model not gravity safe if the scalar potential $V(\phi )$
includes terms which are not suppressed by at least eight 
powers of $1/m_P$~\cite{km_r}. 
The largest dangerous scalar potential terms develop
from interference between $\lambda_2 (XY)^2/m_P$ and 
$\lambda_3 X^pY^q/m_P^{p+q-3}$ when constructing the scalar potential
$V_F=\sum_{\hat{\phi}}|\partial W/\partial\hat{\phi}|_{\hat{\phi}\to\phi}^2$
(here, the $\hat{\phi}$ label chiral superfields with $\phi$ being their leading components).
We find the MBGW model to be not gravity safe under any of the $\mathbb{Z}_N^R$
discrete $R$-symmetries of Table \ref{tab:R}.

%\subsection{Gravity-safety with global symmetry}

%\subsubsection{$n=0$ case:}

%The $n=0$ case has also been examined\cite{Nilles:1981py,Miller:2003hm,
%Ciafaloni:1997gy,Feldstein:2004xi}.
%In this case, the superpotential is given by $W\ni \lambda_\mu XH_uH_d$
%and the model is subject to a PQ global symmetry with $Q_{PQ}(X)=+2$.
%The model lacks the $X^3$ term which is present in the NMSSM.
%The scalar component of $X$ gains a vev $\langle\phi_X\rangle\sim v_{PQ}$
%so that $\lambda_\mu\sim 10^{-9}$. It is argued that stable minima 
%can be found for tiny saxion soft masses and consequently tiny gravitino 
%masses\cite{Feldstein:2004xi}. 
%The cold dark matter is expected to be an axion-quasi-stable saxion admixture
%along with a tiny component of axinos.
%Late-time entropy production is conjectured to occur in the early universe 
%in order to dilute a possible over-abundance of saxions and 
%quasi-stable gravitinos.

\subsection{Hybrid models of PQ breaking from SUSY breaking}
\label{ssec:hybrid}

%\KJB{I redefine $v_{PQ}$ as in the previous subsection and made changes to make things consistent.
%In following plots, we need to change $f_a$ to $v_{PQ}$.}

In this Subsection, we review three models which combine approaches 
where PQ symmetry breaking is triggered by SUSY breaking and 
where a gravity-safe accidental approximate PQ symmetry 
might emerge from a discrete $R$-symmetry.

\begin{itemize}
\item These models are obtained by adopting a hybrid approach~\cite{grav} 
between the radiative breaking models and the MBGW model.
\item In the radiative breaking models, a Majorana neutrino scale is 
generated as the PQ field $X$ gets VEV. However, in the hybrid models, 
the Majorana mass term $MN^cN^c/2$ is allowed but it is not generated through PQ breaking-- 
similar to MBGW model.
\item In the radiative breaking models, intermediate PQ and Majorana neutrino scales
 develop as a consequence of intermediate scale SUSY breaking and the running of 
soft SUSY breaking mass term to negative squared values. In contrast, in the MBGW model
 and in the hybrid models, PQ breaking is triggered by large negative soft terms instead
 of radiative breaking.
\end{itemize}

Three  hybrid models as listed below : 
 
\subsubsection{Hybrid CCK Model}

The superpotential for the hybrid CCK model (hyCCK) is given by~\cite{grav}:
\bea
W_{hyCCK}&\ni &f_uQH_uU^c+f_dQH_d D^c+f_{\ell}LH_dE^c+f_{\nu}LH_uN^c+
M_N N^cN^c/2\nonumber \\
& +& fX^3Y/m_P+\lambda_\mu X^2 H_uH_d/m_P .
\eea
Thus when the PQ symmetry breaks, the $\mu$ parameter is obtained as 

\bea
\mu_{\rm eff} = \lambda_\mu \langle X\rangle^2/m_P .
\eea

We have checked that the hyCCK model is not gravity-safe under the $\mathbb{Z}_N^R$ 
symmetries for $N=4,6,8$ or 12. 
However, it does turns out to be gravity-safe under $\mathbb{Z}_{24}^R$ symmetry
with the $\mathbb{Z}_{24}^R$ charge and PQ charge assignments as shown in Table \ref{tab:hcck}.
\begin{table}[!htb]
\renewcommand{\arraystretch}{1.2}
\begin{center}
\begin{tabular}{c|cccccccccc}
multiplet & $H_u$ & $H_d$ & $Q_i$ & $L_i$ & $U_i^c$ & $D_i^c$ & $E_i^c$ & $N_i^c$ & X & Y \\
\hline
 $\mathbb{Z}_{24}^R$ charge & 16 & 12  & 5 & 9 & 5 & 9 & 5 & 1 & -1 & 5 \\
\hline
PQ charge & -1 & -1  & 1 & 1 & 0 & 0 & 0 & 0 & 1 & -3 \\
\hline
\end{tabular}
\caption{ $\mathbb{Z}_{24}^R$ and PQ charge assignments for various superfields of the 
hyCCK model.
}
\label{tab:hcck}
\end{center}
\end{table} 

\begin{figure}[tbp]
\centering
\includegraphics[height=0.4\textheight]{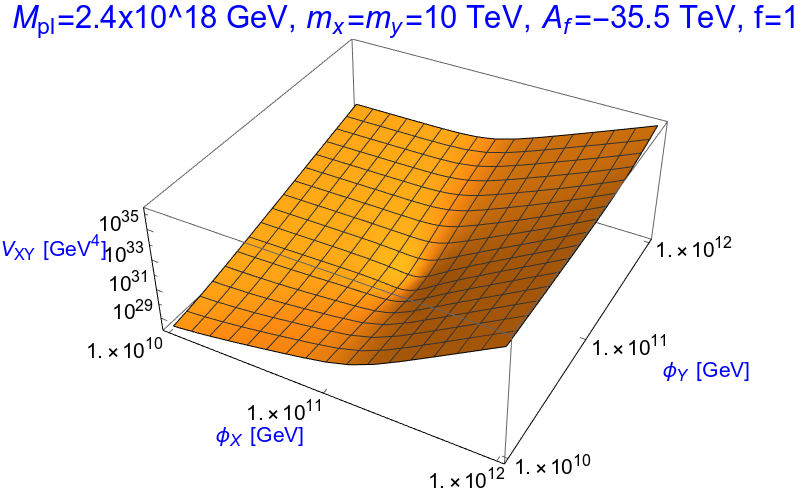}
\caption{Scalar potential $V_{hyCCK}$ versus $\phi_X$ and $\phi_Y$ for
$m_X=m_Y\equiv m_{3/2}=10$ TeV, $f=1$ and $A_f=-35.5$ TeV.
\label{fig:Vgspq}}
\end{figure}

The scalar potential for hyCCK is found to be
\bea
    V = [fA_f\frac{\phi_X^3\phi_Y}{m_P}+h.c.] + m_X^2|\phi_X|^2 + m_Y^2|\phi_Y|^2 + \frac{f^2}{m_P^2}[9|\phi_X|^4|\phi_Y|^2 + |\phi_X|^6]
\eea
and is shown in Fig. \ref{fig:Vgspq} vs. scalar field values $\phi_X$ and $\phi_Y$. For large
negative values of soft term $A_f$, then a $\mathbb{Z}_{24}^R$ and $PQ$ breaking minimum develops.

The lowest order PQ violating terms in the superpotential are 
$X^8Y^2/ m_P^7$, $X^4Y^6/m_P^7$ and $Y^{10}/ m_P^7$ which implies that the lowest order 
PQ breaking term in the scalar potential is suppressed by $1/ m_P^8$. 
Therefore, this model satisfies the KMR condition for being gravity-safe.

\begin{figure}[tbp]
\centering
\includegraphics[height=0.4\textheight]{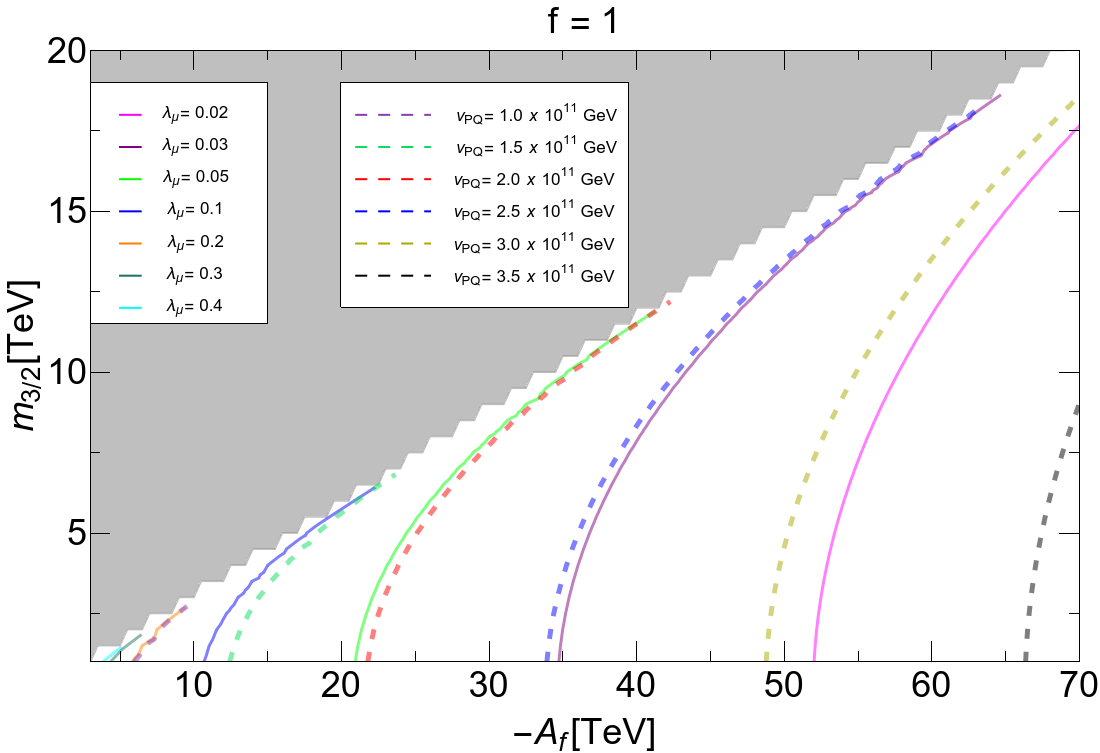}
\caption{Representative values of $\lambda_{\mu}$ required for $\mu =200$ GeV 
in the $m_{3/2}$ vs. $-A_f$ plane of the hyCCK model for $f=1$. 
We also show several contours of $v_{PQ}$.
%\KJB{Text says $\mu=150$ GeV, please check it.}
\label{fig:GSPQ}}
\end{figure}

The allowed range of hyCCK model parameter space is shown in Fig. \ref{fig:GSPQ}
where we show contours of $\lambda_{\mu}$ values which lead to $\mu =200$ GeV
in the $m_{3/2}$ vs. $-A_f$ plane for $f=1$. 
We also show several representative contours of $v_{PQ}$ values.
Values of $\lambda_{\mu}\sim 0.015-0.2$ are generally sufficient for a natural
$\mu$ term and are easily consistent with soft mass 
$m_{soft}\sim m_{3/2}\sim 2-30$ TeV as indicated by LHC searches.
We also note that for $m_{3/2}\sim 5-20$ TeV, then $v_{PQ}\sim 10^{11}$ GeV which corresponds 
to the sweet spot for axion cold dark matter. 

\subsubsection{Hybrid SPM Model}

The superpotential for the hybrid SPM model (hySPM) is given by~\cite{Choi:2010xf,grav}
\bea
W_{hySPM}&\ni &f_uQH_uU^c+f_dQH_d D^c+f_{\ell}LH_dE^c+f_{\nu}LH_uN^c+M_NN^cN^c/2\nonumber \\
& +& fX^3Y/m_P+\lambda_\mu Y^2 H_uH_d/m_P  .
\eea
In this case, when PQ symmetry breaks, the $\mu$ parameter is generated to be
\bea
\mu_{\rm eff} = \lambda_\mu \langle Y\rangle^2/m_P .
\eea
This model also turns out to be not gravity-safe under $\mathbb{Z}_N^R$ symmetries for 
$N=4,6,8$ and 12 but is gravity-safe for $\mathbb{Z}_{24}^R$ symmetry.
The gravity-safe $\mathbb{Z}_{24}^R$ charge 
and PQ charge assignments are shown in Table \ref{tab:hspm}.
\begin{table}[!htb]
\renewcommand{\arraystretch}{1.2}
\begin{center}
\begin{tabular}{c|cccccccccc}
multiplet & $H_u$ & $H_d$ & $Q_i$ & $L_i$ & $U_i^c$ & $D_i^c$ & $E_i^c$ & $N_i^c$ & X & Y \\
\hline
 $\mathbb{Z}_{24}^R$ charge & 16 & 12  & 5 & 9 & 5 & 9 & 5 & 1 & 5 & -13 \\
\hline
PQ charge & -1 & -1  & 1 & 1 & 0 & 0 & 0 & 0 & -1/3 & 1 \\
\hline
\end{tabular}
\caption{ $\mathbb{Z}_{24}^R$ and PQ charge assignments for various superfields of the 
hySPM model.
}
\label{tab:hspm}
\end{center}
\end{table} 

The scalar potential is obtained similar to that in the hyCCK model with the 
only difference being that now the lowest order  PQ violating terms in the superpotential are 
$Y^8X^2/m_P^7$, $Y^4X^6/ m_P^7$ and $X^{10}/m_P^7$ which means that the lowest order PQ breaking terms 
in the scalar potential are suppressed by $1/ m_P^8$ so that the hySPM model also satisfies
the KMR condition for being gravity-safe.  

\begin{figure}[tbp]
\centering
\includegraphics[height=0.4\textheight]{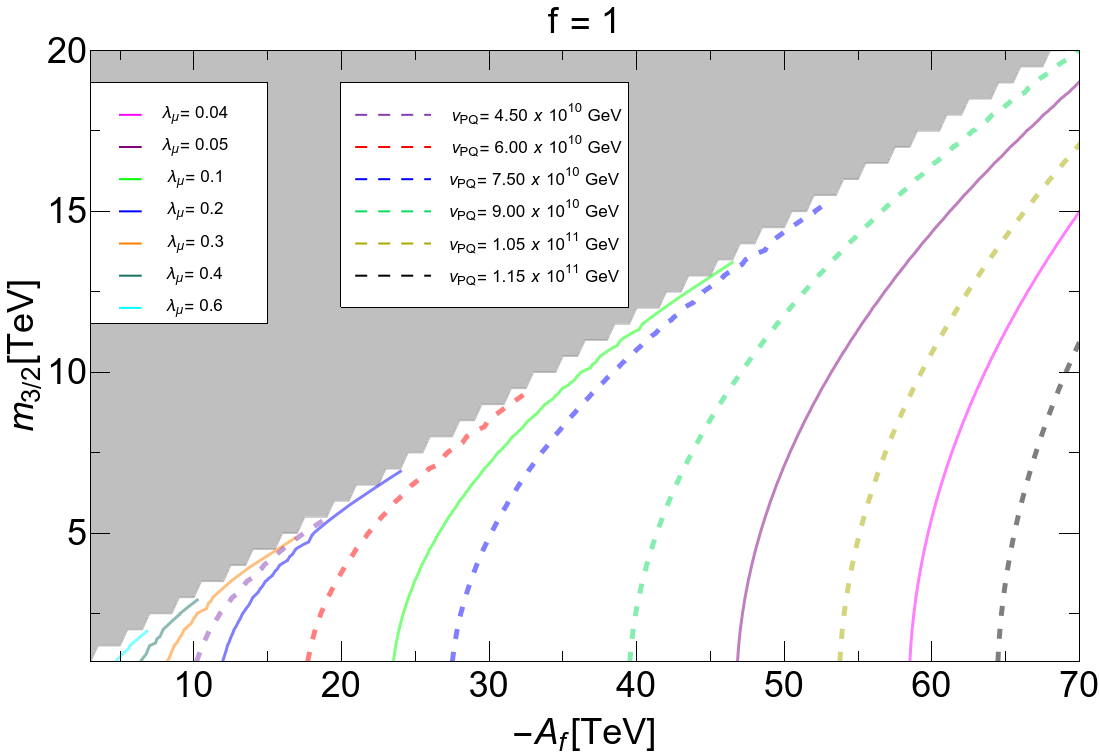}
\caption{Representative values of $\lambda_{\mu}$ required for $\mu =150$ GeV in the 
$m_{3/2}$ vs. $-A_f$ plane of the hySPM model for $f=1$. 
We also show several contours of $v_{PQ}$.
\label{fig:spm}}
\end{figure}

The allowed range of hySPM model parameter space is shown in Fig. \ref{fig:spm}
where we show contours of $\lambda_{\mu}$ values which lead to $\mu =150$ GeV
in the $m_{3/2}$ vs. $-A_f$ plane for $f=1$. 
We also show several representative contours of $v_{PQ}$ values.

\subsubsection{Hybrid MSY model}

The superpotential in the hybrid MSY model (hyMSY) is given as~\cite{grav}:
\bea
W_{hyMSY}&\ni &f_uQH_uU^c+f_dQH_d D^c+f_{\ell}LH_dE^c+f_{\nu}LH_uN^c+M_NN^cN^c/2\nonumber \\
& +& fX^3Y/m_P+\lambda_\mu XY H_uH_d/m_P .
\eea
However, we have checked that the hyMSY model does not satisfy
the KMR condition for being gravity-safe 
under any of the $R$-symmetries listed in Table \ref{tab:R}.

\section{Are the various $\mu$ solutions experimentally 
distinguishable?}
\label{sec:exp}

An important question arises: are the various solutions to the SUSY $\mu$ problem 
experimentally testable and experimentally distinguishable from one another? 

Obviously, one important consequence is the existence of weak scale SUSY (WSS) 
so that if WSS is disproved, then the whole discussion on the origin of the $\mu$ term is moot.
The main {\it raison d'etre} for SUSY is to stabilize the weak scale under the presence 
of quantum corrections. 
In addition, WSS provides a natural mechanism for electroweak symmetry breaking.
This means no severe fine-tuning of parameters involved in determining
the magnitude of the weak scale, which we take to be no  fine-tuning in Eq. \eqref{eq:mzs}.
Upper limits have been derived on sparticle masses within the context of unified SUSY 
models with no fine-tuning~\cite{rns2,upper,Baer:2018hpb} ({\it i.e.} $\Delta_{EW}\alt 30$). 
These imply typically $m_{\tg}\alt 6$ TeV and $m_{\tst_1}\alt 3$ TeV and $|\mu |\alt 360$ GeV.
To explore such high sparticle masses, then about 15 ab$^{-1}$ of $pp$ collisions
at $\sqrt{s}\agt 27$ TeV is required for a hadron collider~\cite{Baer:2018hpb} or
$\sqrt{s}\agt 720$ GeV is needed for an $e^+e^-$ collider~\cite{Baer:2014yta}. 
If no sparticles are seen at such colliders, 
then SUSY as we understand it would no longer be a viable hypothesis for 
stabilization of the weak scale.

Some of the $\mu$ solutions are expected to give rise to the MSSM-only as the weak scale
effective theory. In this case, it may be difficult to distinguish for instance a
GM solution from a CM solution. In the case of the G2MSSM solution, distinctive 
mass relations amongst sparticles are expected to occur which could support
or deter such explanations~\cite{Acharya:2012tw}.

In addition to weak scale SUSY, several models-- KN, CKN, EWK, HFD,
radiative PQ models (MSY, CCK, SPM),
MBGW and hybrid models predict a SUSY DFSZ axion. 
Recent searches for axions at axion haloscope experiments~\cite{admx} have reached the 
non-SUSY DFSZ coupling strengths for a narrow range of $m_a$ possibilities.
However, the SUSY DFSZ axion-- by virtue of including higgsinos in the $a\gamma\gamma$ triangle
vertex-- has a much smaller coupling~\cite{axpaper}. 
It is not clear whether present technology has the 
capability to probe such tiny $a\gamma\gamma$ couplings. In the event that a thorough
search can be made for SUSY DFSZ axions over their allowed range of masses and couplings 
strengths, then (non)observation of axions could rule out or verify this class of $\mu$
problem solutions. A related test could be the determination of a diminished abundance
of higgsino-like WIMPs such that the presence (or not) of additional dark matter particles 
such as axions is required.

Several of the $\mu$ solutions require as well additional distinctive particles.
The NMSSM solution requires the presence of additional scalar and pseudoscalar Higgs bosons
and a fifth neutralino arising from the NMSSM singlino. For many NMSSM parameter choices,
some deviations in the $h$ boson coupling strengths are expected~\cite{nmssm,balazs}.

The $U(1)^\prime$ $\mu$ solutions also include distinctive new particle predictions. 
The CDEEL model~\cite{cvetic} requires the presence of an additional weak scale $Z^\prime$ boson
which could decay to higgsinos as well as SM particles~\cite{arvanitaki,cp}. 
For the HPT model~\cite{xt}, the $Z^\prime$ is expected to be far beyond any collider reach projections. 
Instead, for HPT, 
one expects bilinear RPV leading to distinctive collider signatures and altered
expectations for dark matter. Also, in these models one may expect the presence of
stable weak scale exotic hadrons or other exotica which arise from the requirement for
anomaly cancellation.

\section{Conclusions} 
\label{sec:conclude}

In this paper, we have re-examined the SUSY $\mu$ problem with
perspective gained from experimental results from LHC through Run 2
with 150 fb$^{-1}$ of data. The two parts to the SUSY $\mu$ solutions are 
1. first forbid the $\mu$ term, perhaps via some symmetry and then 
2. regenerate it, perhaps via symmetry breaking. The new perspective from
LHC and the naturalness issue is that $\mu$ should be generated
of order $m_{weak}\sim m_{W,Z,h}\sim 100-300$ GeV whilst the soft SUSY breaking
terms likely inhabit the multi-TeV regime. Thus, a Little Hierarchy (LH)
should now be included in SUSY $\mu$ solutions where $|\mu |\ll m_{soft}$.
This is different from pre-LHC expectations where solutions
sought to generate $|\mu |\simeq m_{soft}$.

To gain an updated perspective on the SUSY $\mu$ problem, we examined
twenty solutions. 
These solutions are summarized in Table \ref{tab:overview} where we list 
each solution and how it may admit a LH, whether it also addresses 
the strong CP problem, whether it is gravity-safe, 
its relation to neutrino masses
(Standard see-saw or other) and any distinctive experimental consequences.
While all solutions have the capacity to be consistent with the LH (usually
by adjusting some arbitrary constant $\lambda_{\mu}$), some actually
generate $\mu\sim m_{weak}\ll m_{soft}$ with $\lambda_\mu\sim 1$ (such as the
radiative PQ breaking models MSY, CCK and SPM). 

Also, early attempts to solve the SUSY $\mu$ problem could appeal to an
underlying global symmetry such as PQ to suppress the $\mu$ term. 
It soon became clear that such global symmetries are not consistent with
an ultra-violet completion which includes gravity effects since gravitational
interactions don't respect global symmetries. Continuous ($U(1)^\prime$) or
discrete gauge symmetries are gravity-safe but usually require the addition
of perhaps unwanted exotica in order to preserve anomaly-freedom. 
The more recent emergence of discrete $R$-symmetries~\cite{lrrrssv1,lrrrssv2}, 
which can arise 
from compactification of extra dimensions in string theory, seems to 
provide the cleanest suppression symmetry for the $\mu$ term.
A delineation of anomaly-free (including a GS term) 
$\mathbb{Z}_N^R$ symmetries which are consistent with $SO(10)$ or $SU(5)$
unification (thus preserving gauge coupling unification) offers perhaps
the most compelling solutions for the first half of the SUSY $\mu$ problem.
For $N=4,6,8,12$ and 24, these symmetries forbid $\mu$ along with RPV
trilinear terms and dimension-5 $p$-decay operators whilst allowing
the required Yukawa couplings and neutrino mass operators. Of these, 
the $\mathbb{Z}_4^R$ stands out as both simple and compelling.
It should probably now replace $R$-parity as a standard pillar 
upon which the MSSM is constructed. 

If one also seeks to simultaneously solve the strong CP problem, 
then the $\mathbb{Z}_{24}^R$ symmetry works in the hybrid models to suppress
unwanted superpotential terms while providing the underlying fundamental
symmetry from which a global PQ can emerge as an accidental, approximate
symmetry which is gravity-safe.
Several other solutions also have their roots in stringy behavior
(CM, $U(1)^\prime$, instanton, G2MSSM).

If the naturalness edict is followed-- which requires $|\mu |$ not too far
from $m_{weak}\sim 100$ GeV-- then one expects thermally-underproduced
higgsino-like WIMPs as (part of) dark matter. If the natural WIMP
abundance is enhanced by non-thermal processes to make up the 
entirety of dark matter, then they become excluded by a combination
of direct and indirect WIMP detection experiments~\cite{Baer:2018rhs}. 
Thus, additional dark matter beyond WIMPs then seems to be required. 
The axion is a highly motivated candidate to make up the remaining 
bulk of dark matter. To gain accord with the requirements of 
cold dark matter, a gravity-safe solution to the strong CP problem
and a solution to the SUSY $\mu$ problem (while also suppressing dangerous 
$p$-decay operators and allowing for see-saw neutrino masses), then the
hybrid models based on $\mathbb{Z}_{24}^R$ discrete $R$-symmetry stand out
as a rather complete answer.

Overall, the SUSY $\mu$ problem has generated a rich panoply of 
solutions over the past 35 years. To begin the process of selecting amongst 
them or building others, it is of the essence to first discover SUSY
and then to proceed with precision measurements of the SUSY spectra along with
any exotica to gain insight into which if any of the solutions 
best describes nature.
Future collider and dark matter experiments should go a long way towards
selecting amongst or ruling out these various solutions and 
other solutions perhaps yet to come.

\begin{table}[!htb]
\renewcommand{\arraystretch}{1.2}
\begin{center}
\begin{tabular}{|c|ccccc|}
\hline
model & admit LH? & strong CP? & gravity safe? & see-saw? & exp. cons.\\
\hline
GM & small $\lambda_{\mu}$ & $\times$ & $--$ & $SNSS$ & MSSM \\
\hline
CM & small $\lambda_{\mu}$ & $\times$ & $--$ & $SNSS$ & MSSM \\
\hline
$R$-sym & $(v_i/m_P)^{n_i}\ll 1$ & $\times$ & $?$ & $SNSS$ & MSSM \\
\hline
$\mathbb{Z}_4^R$ & small $\lambda_{\mu}$ & $\times$ & $--$ & $SNSS$ & MSSM \\
\hline
Instanton & small $e^{-S_{cl}}$ & $\times$ & $--$ & $SNSS$ & MSSM \\
\hline
$G_2MSSM$ & $\langle S_i\rangle/m_P\ll 1$ & $\times$ & $--$ & $SNSS$ & $G_2MSSM$ \\
\hline
NMSSM & small $\lambda_{\mu}$ & $\times$ & $--$ & $SNSS$ & extra Higgs/neutralino\\
\hline
nMSSM & small $\lambda_{\mu}$ & $\times$ & $--$ & $SNSS$ & extra Higgs/neutralino\\
\hline
$\mu\nu$SSM & small $\lambda_{\mu}$ & $\times$ & $--$ & $bRPV$ & $bRPV$, mixings\\
\hline
$U(1)^\prime$ (CDEEL)  & small $\lambda_{\mu}$ & $\times$ & $--$ & $SNSS$ & $Z^\prime$ \\
\hline
sMSSM & small $\lambda_{\mu}$ & $\times$ & $--$ & $SNSS$ & extra Higgs/neutralino\\
\hline
$U(1)^\prime$ (HPT) & small $\lambda_{\mu}$ & $\times$ & $--$ & $bRPV$ 
& $bRPV$, stable heavy hadrons \\
\hline
KN & $v_{PQ}<m_{hidden}$ & $\surd$ & $?$ & $SNSS$ & DFSZ axion\\
\hline
CKN & $\Lambda <\Lambda_h$ & $\surd$ & $?$ & $SNSS$ & DFSZ axion\\
\hline
BK/EWK & $\lambda_\mu\sim 10^{-10}$ & $\surd$ & $?$ & $SNSS$ & DFSZ axion\\
\hline
$\rm HFD$ & $v_{PQ}< m_{hidden}$ & $\surd$ & $?$ & $SNSS$ & MSSM \\
\hline
MSY/CCK/SPM & $v_{PQ}< m_{hidden}$ & $\surd$ & $\times$ & $RadSS$ & DFSZ axion \\
\hline
CCL & small $\lambda_{\mu}$ & $\surd$ & $?$ & $several$ & DFSZ axion, $\tG$ or $\tnu$ LSP \\
\hline
BGW & small $\lambda_{\mu}$ & $\surd$ & $\mathbb{Z}_{22}$ & $SNSS$ & DFSZ axion \\
\hline
Hybrid CCK/SPM & small $\lambda_{\mu}$ & $\surd$ & $\mathbb{Z}_{24}^R$ & $SNSS$ & DFSZ axion \\
\hline
\end{tabular}
\caption{Summary of twenty solutions to the SUSY $\mu$ problem and how they
1. admit a Little Hierarchy (LH), 2. solve the strong CP problem 
($\surd$) or not ($\times$), 
3. are expected gravity-safe, 
4. Standard neutrino see-saw (SNSS) or other and 
5. some experimental consequences.
}
\label{tab:overview}
\end{center}
\end{table}

\section*{Acknowledgments}

We thank H. Serce for help in the early stages of this project.
This work was supported in part by the US Department of Energy, Office of High Energy Physics.
The work of KJB was supported by IBS under the project code, IBS-R018-D1.

%
%%%%%%%%%%%%%%%%%%%%%%%%%%%%%%%%%%%%%%%%%%%%%%%%%%%%%%

%
%
\end{document}